\documentclass[journal]{IEEEtran}
\ifCLASSINFOpdf
\else
\fi

\usepackage{algorithm}
\usepackage{algorithmicx}
\usepackage{algpseudocode}
\usepackage{amsmath}
\usepackage{multirow}
\usepackage[normalem]{ulem}
\useunder{\uline}{\ul}{}

\usepackage{multirow}
\usepackage{ dsfont }
\usepackage{graphicx}
\usepackage{caption}
\usepackage{subfigure}
\usepackage[table,xcdraw]{xcolor}
\usepackage{orcidlink}

\begin{document}

\title{Hybrid Reinforcement Learning-Based Eco-Driving Strategy for Connected and Automated Vehicles at Signalized Intersections}

\author{Zhengwei~Bai$^{\orcidlink{0000-0002-4867-021X}}$,~\IEEEmembership{Student Member,~IEEE},
        Peng~Hao,~\IEEEmembership{Member,~IEEE},
        Wei Shangguan,~\IEEEmembership{Member,~IEEE},\\
        Baigen Cai,~\IEEEmembership{Senior Member,~IEEE},
        and~Matthew~J.~Barth$^{\orcidlink{0000-0002-4735-5859}}$,~\IEEEmembership{Fellow,~IEEE}
\thanks{Zhengwei Bai and Matthew J. Barth are with the Department of Electrical and Computer Engineering, University of California at Riverside, Riverside, CA 92507 USA (e-mail: zbai012@ucr.edu).}
\thanks{Peng Hao is with the Center for Environmental Research
and Technology, University of California at Riverside, Riverside, CA 92507 USA.}
\thanks{Wei Shangguan and Baigen Cai are with the State Key Laboratory of Rail Traffic Control and Safety and the Beijing Engineering Research Center of EMC and GNSS Technology for Rail Transportation, School of Electronic and Information Engineering, Beijing Jiaotong University, Beijing 100044, China.}

}

\maketitle

\begin{abstract}
Taking advantage of both vehicle-to-everything (V2X) communication and automated driving technology, connected and automated vehicles are quickly becoming one of the transformative solutions to many transportation problems. However, in a mixed traffic environment at signalized intersections, it is still a challenging task to improve overall throughput and energy efficiency considering the complexity and uncertainty in the traffic system. In this study, we proposed a hybrid reinforcement learning (HRL) framework which combines the rule-based strategy and the deep reinforcement learning (deep RL) to support connected eco-driving at signalized intersections in mixed traffic. Vision-perceptive methods are integrated with vehicle-to-infrastructure (V2I) communications to achieve higher mobility and energy efficiency in mixed connected traffic. The HRL framework has three components: a rule-based driving manager that operates the collaboration between the rule-based policies and the RL policy; a multi-stream neural network that extracts the hidden features of vision and V2I information; and a deep RL-based policy network that generate both longitudinal and lateral eco-driving actions. In order to evaluate our approach, we developed a Unity-based simulator and designed a mixed-traffic intersection scenario. Moreover, several baselines were implemented to compare with our new design, and numerical experiments were conducted to test the performance of the HRL model. The experiments show that our HRL method can reduce energy consumption by 12.70\% and save 11.75\% travel time when compared with a state-of-the-art model-based Eco-Driving approach.

\end{abstract}


%
\IEEEpeerreviewmaketitle

\section{Introduction}
%
%
%
%
\IEEEPARstart{W}{ith} the rapid development of vehicle communication technology, connected vehicles (CVs) have shown the capability to improve traffic mobility and energy efficiency via vehicle-to-vehicle (V2V) or vehicle-to-infrastructure (V2I) communication \cite{vahidi2018energy}. Meanwhile, automated vehicles (AVs) equipped with sensing technology (e.g. camera, Lidar, radar,  etc.) and artificial intelligent (AI) technology can recognize the environment and subsequently perform proper actions by fully or partial automation \cite{fagnant2015preparing}. Taking advantage of both vehicle-to-everything (V2X) communication and autonomous driving technology, connected and automated vehicles (CAVs) have emerged as one of the transformative solutions to the current challenges in transportation, such as traffic congestion, air pollution, and energy consumption \cite{INRIX,GGE}. Urban traffic at signalized intersections is one good scenario to utilize the advantages of CV-AV fusion, as the vehicles may have extensive interaction with traffic signal timing and non-surrounding vehicles, which is imperceptible from on-board sensors but detectable via V2X communications.

In order to improve traffic efficiency and energy savings around intersections, a series of CAV-based studies have been conducted. Rios-Torres et al. summarized the relative studies and the research trend of CAVs on the connectivity-based intersection driving optimization \cite{rios2016survey}. Lee et al. proposed a cooperative vehicle intersection control (CVIC) system to enable the interaction between vehicles and infrastructure to optimize the efficient intersection operation \cite{lee2012development}. According to the study conducted by Guler et al., the increase in the penetration rate of CV in low demand traffic can significantly reduce the average delay of passing intersections by applying connected platooning control \cite{guler2014using}. Elhenawy et al. proposed a game-theory-based control algorithm for cooperative adaptive cruise control (CACC) based AVs, which reduces travel time and delay \cite{elhenawy2015intersection}.  From the eco-driving perspective, De Nunzio et al. proposed an optimal-control-based eco-driving algorithm under traffic-free (i.e. there are no other vehicles) intersection networks by using V2I-based traffic signal data \cite{zhang2016optimal}. Hao et al. proposed an Eco-Approach and Departure (EAD) system and evaluated the EAD application in real-world traffic, which takes the advantage of signal phase and timing (SPaT) and Geometric Intersection Description (GID) information broadcasted by the traffic signals \cite{hao2018eco}.  Fei et al. proposed an eco-driving strategy by predicting the speed of the preceding vehicle, which can work well under congested intersections \cite{ye2018prediction}. In addition, according to the research conducted by Wang et al., the penetration rate of CAVs positively affects the energy efficiency of the signalized traffic network \cite{wang2019cooperative}. The aforementioned approaches are, to the most extent, rule-based or model-based algorithms. These traditional rule-based and model-based driving control strategy can generate precise control trajectories and is easy to define and recognize, but most of these conventional algorithms are dependent on ideal assumptions. For instance, some intersection-cooperative driving methods \cite{de2016eco,xia2012field} assume that the environment is traffic-free, which may fail to work in real-world traffic conditions. Some studies (e.g., \cite{zhang2016optimal,lee2012development}) assume 100\% CV penetration rate (i.e. all vehicles are fully connected), which is not realistic at least in the near future. Additionally, some eco-driving approaches \cite{hao2018connected,hao2018eco,ye2017prediction} only care about longitudinal maneuver to improve the energy performance, ignoring the potential of lateral maneuvers under a dynamic traffic environment. These simplifications of the real system limit the capability to incorporate complex dynamics when the system environment is complex or unknown. 
    
Hence, to overcome the of longitudinal-only maneuvers, Jia Hu et al. proposed an optimization-based overtaking method for eco-approach under a signalized intersection \cite{hu2021cut}. The eco-approach process is formulated into an optimal problem constrained by the energy-saving purpose. Considering the cooperation between CAVs, AVs and Human-driven Vehicles (HVs), Weiming Zhao et al. proposed a  platoon based cooperative eco-driving model under mixed traffic environment based on model predictive control (MPC) method. The results show that the proposed model does not compromise the traffic efficiency and the driving comfort while achieving the eco-driving strategy \cite{zhao2018platoon}. Huang et al. \cite{huang2018eco} provided a detailed review on model-based Eco-Driving methods and discussed the limitations and future directions of the Eco-Driving tasks. In summary, the main advantages of the model-based Eco-Driving approaches are evident, e.g., stable performance, interpretative mechanism, and readily for commercialized implementation. Nevertheless, the uncertainty of traffic environment, along with the complexity of the driving strategies will significantly increase the difficult for designing such a model-based methods, thus diminishing their advantages. Currently, most of the model-based Eco-Driving approaches usually integrate rule-based schemes to decompose the overall Eco-Driving task into several sub-tasks that can be solved individually. Hence the generalization and adaptation ability of these methods can be further improved.

Recently, with the development of machine learning (ML) \cite{jordan2015machine}, it is possible to solve the aforementioned constrains by utilizing the tremendous generalization power of deep learning (DL) \cite{lecun2015deep} and reinforcement learning (RL) \cite{sutton2018reinforcement}, which do not rely on specific models or rules. Particularly, reinforcement learning has demonstrated its significant power in dealing with policy learning tasks by itself without predefined human rules or models in a complex environment \cite{Silver2016Mastering,Mnih2015Human}, including transportation. Owing to the great capability of RL, many researchers are trying to apply RL algorithms into autonomous driving tasks. A deep RL-based autonomous driving framework was proposed by Sallab et al. to enable automatic lane-keeping with interaction with simple traffic \cite{sallab2017deep}. Desjardins et al. proposed an RL-based cooperative adaptive cruise control (CACC) method by utilizing V2V information, which can result in efficient behavior in CACC \cite{desjardins2011cooperative}. Shalev-Shwartz proposed an RL-based safe driving model, which enables multi-agents to merge smoothly in a double-merge scenario \cite{shalev2016safe} . For signalized intersection scenario, Chen et al. proposed a hybrid reinforcement learning based driving behavior control model, which enables the vehicle to interact with the traffic signal (i.e. stop or go with different phase) \cite{chen2018deep} . Most existing RL-based algorithms focused on lane-keeping, CACC, merging and traffic-signal interaction, but few RL algorithms are used in long-term intersection-based eco-driving (LIED) strategy to the authors' knowledge. One reason may be that RL algorithms are good at solving single logical tasks, while the LIED is a multi-logical task with at least three different tasks: (1) Energy-efficient driving, which requires the vehicle to drive through the intersection with less energy consumption. (2) Interaction with signals, which requires the vehicle to react properly with the traffic signal; and (3) Interaction with traffic, which requires the vehicle to keep safe distance with other vehicles.

In this study, to further explore the eco-driving strategy of CAV under mixed connected traffic around a signalized intersection, we proposed a hybrid reinforcement learning (HRL) framework to learn the LIED strategies. The key innovations of this research include: (1) we propose an HRL framework for a logically complex task, i.e. time-efficient eco-driving in signalized intersections with mixed traffic; (2) an innovative long-short term reward (LSTR) model to provide the RL model with the ability to learn complex driving strategies from conflicting factors (i.e. speeding-up and energy-saving); (3) The traffic environment is considered as mixed traffic where the other vehicles are human-driven and unconnected with diverse dynamic characteristics; (4) a multi-sensor-based RL-network is used to enable ego-vehicles to interact properly with the mixed traffic; and (5) a game-engine based simulation platform is designed and numerical experiments are conducted to assess the proposed model compared with several baselines.

The following section provides details of the problem formulation, which includes the design of traffic environment and ego-vehicle, along with the HRL-based eco-driving framework, which includes the system architecture, algorithms design of decision manager, preprocessing, eco-driving RL algorithm, and the definition of network structure. The methodology section is followed by the experiment section which introduces the simulator development and the procedures of training and testing. Next, the Game-Engine based simulator design and comprehensive numerical experiments are conducted to comprehensively assess the performance of the proposed method. The last section concludes the paper with a discussion on future work.

\section{Problem Formulation}
The main purpose of this study is to design an RL-based framework for vision-based CAVs to perform eco-driving strategies under mixed traffic in the signalized intersection. This topic is substantially an optimal policy learning task. There are three main goals of the policy: (1) to save energy, (2) to reduce the travel time, and (3) to interact with traffic and signals safely. 

RL has five key compositions \cite{1998Reinforcement} and the connections between these compositions and the traffic elements in this paper are shown as follows:
\begin{itemize}
\item Environment: traffic environment which includes various kinds of vehicles and a fixed-timing signalized intersection;
\item Agent: the ego-vehicle which can perceive the environment via a front camera and V2V-based SPaT (Signal Phase and Timing) information;
\item Policy: the proposed eco-driving policy;
\item Action reward: the short-term benefit (i.e. speed reward, energy consumption) of taking action right at this moment and the long-term benefit (i.e. total travel time, total energy consumption) of the journey;
\item Action-value function: the function to determine which action is the best choice at the next moment to achieve a long-term optimal result.
\end{itemize}
    
The above elements indicate that the issue in this research can be well interpreted by the RL framework. The RL framework is established based on Markov Decision Process (MDP) which is a mathematical framework for decision making via the interaction between a learning agent (the ego-vehicle in this paper) and its environment (i.e. mixed traffic and signalized intersection) in terms of state, actions and rewards \cite{bertsekas1995dynamic}.
    
To be specific, the mixed traffic environment at a signalized intersection (Environment) and HRL-based ego-vehicle (Agent) are discussed below in this section. Additionally, the HRL framework (Policy), the LSTR model (Action reward), and Deep RL Network (Action-value function) are discussed in the next section, respectively.

\subsection{Traffic Environment}

\begin{figure}
    \centering
    \includegraphics[width=0.47\textwidth]{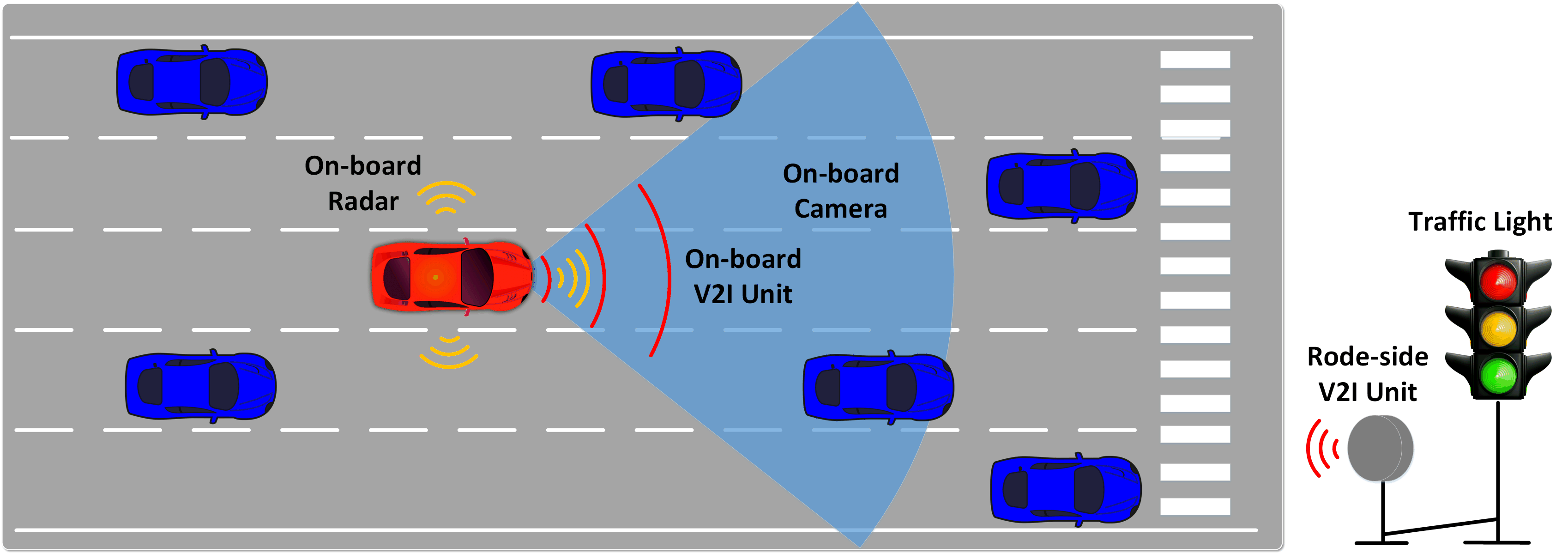}
    \caption{The traffic environment visualization which includes ego-vehicle (red), other vehicles (blue), traffic signal, road-side V2I unit, and on-board sensors. }
    \label{fig:environment}
\end{figure}

The traffic environment includes three main components: the ego-vehicle, conventional vehicles (i.e. other vehicles in the simulation), and a five-lane signalized intersection (as an example), which is shown as Figure~\ref{fig:environment}. 

To make the proposed environment similar to  real traffic, we design various vehicle types and dynamic traffic signal scenarios. To be specific, vehicles in the simulation consist of five different dynamic models and behavior strategies. For the conventional vehicles' longitudinal control, we applied intelligent driver model (IDM) \cite{treiber2000congested} to generate the instant longitudinal acceleration, which is shown as follows:
\begin{equation}
a_{free}^{(i)}(t)=a \cdot \left(1- \left(\frac{v(t)}{v_{tar}}\right)^{2}\right)
\end{equation}

\begin{equation}
a_{int}^{(i)}(t)=a \cdot \left[1-\left(\frac{v(t)}{v_{tar}}\right)^{4} - \left(\frac{s_{0} + v(t)T + \frac{v(t)\Delta v}{2\sqrt{a b}}}{s}\right)^{2}\right]
\end{equation}

\begin{equation}
\Ddot{x}_{lon}^{(i)}(t)=\left\{
\begin{array}{rcl}
a_{free}^{(i)}(t), &\text{no leading vehicle}\\
a_{int}^{(i)}(t), & \text{otherwise}
\end{array} \right.
\end{equation}
where $a_{free}^{(i)}(t)$ and $a_{int}^{(i)}(t)$ represent the \textit{i}-th vehicle's acceleration when there is no leading vehicle and there is a leading vehicle in the current lane, respectively. Specifically, parameters $a, b, v, v_{tar}, T, \Delta v$, and $s_{0}$ represent the maximum acceleration, maximum deceleration, current longitudinal speed, target speed, safe time headway, speed difference with front vehicle, and minimum distance for the \textit{i}-th vehicle respectively. Finally, $\Ddot{x}_{lon}^{(i)}(t)$ represents the longitudinal acceleration for the vehicle.

For the lateral control, we designed various lane changing rates, $R_{lat}$, for conventional vehicles to describe the probability for taking the lane-changing maneuvers at each time step $t$. Furthermore, this $R_{lat}$ can also represent the human drivers' lateral driving behaviors. In this paper, the low-level lateral control models are designed as follows:
\begin{equation}
x_{lat}^{(i)}(t+1)=\left\{
\begin{array}{rcl}
x_{lat}^{(i)}(t)+v_{lat}, & r_{lat}(t) \leq \dfrac{R_{lat}^{(i)}}{2} \\
x_{lat}^{(i)}(t)\quad\quad\quad, & \dfrac{R_{lat}^{(i)}}{2} < r_{lat}(t) \leq 1-\dfrac{R_{lat}}{2}\\
x_{lat}^{(i)}(t)-v_{lat}, & r_{lat}(t) > 1-\dfrac{R_{lat}^{(i)}}{2}
\end{array} \right.
\end{equation}
where $r_{lat}(t)$, $x_{lat}^{(i)}(t)$, $v_{lat}$ represent the random lane-changing flag, lateral position and lateral driving velocity, for the \textit{i}-th vehicle at time step \textit{t}, respectively. The $r_{lat}(t)\sim U(0, 1)$ is updated at each time step $t$. Table \ref{tab:carModel} shows more details about the kinematic model for the various conventional vehicles.

\begin{table}[ht]
    \centering
    \caption{The description of the kinematic model of the conventional vehicles.}
    \label{tab:carModel}
    \begin{tabular}{c|c|c|c|c|c|c}
    \hline
    \textbf{Vehicle} & \textbf{$a$} & \textbf{$b$} & \textbf{$s_{0}$}   & \textbf{$T$} & \textbf{$v_{tar}$} & \textbf{$R_{lat}$} \\\hline
    \textbf{Type A} & $6.0m/s^{2}$  &$6.0m/s^{2}$ & $3m$ & $1.5s$ & $13.8m/s$ & 0.3 \\
    \textbf{Type B} & $5.0m/s^{2}$  &$4.5m/s^{2}$ & $3m$ & $1.5s$ & $12.5m/s$ & 0.2 \\
    \textbf{Type C} & $3.0m/s^{2}$  &$5.0m/s^{2}$ & $2m$ & $1.2s$ & $11.1m/s$ & 0.2 \\
    \textbf{Type D} & $3.0m/s^{2}$  &$3.0m/s^{2}$ & $3m$ & $1.5s$ & $9.72m/s$ & 0.1 \\
    \textbf{Type F} & $2.0m/s^{2}$  &$1.5m/s^{2}$ & $5m$ & $1.5s$ & $8.33m/s$ & 0.1 \\\hline
    \end{tabular}
    
\end{table}
        
For the traffic signal, we used fixed timing plan with 20s, 3s, 40s, and 1s as the duration of green time, yellow time, red time and all-red time phases respectively. Every time the simulation starts, the initial phase and time are randomly sampled in the entire cycle duration. 

\subsection{Agent}

In this paper, the ego-vehicle is modeled as an electric CAV. The maximum acceleration and deceleration are set as $3m/s^{2}$ and $3m/s^{2}$ respectively. Furthermore, the input observation, energy consumption model (ECM) and output action are defined as follows:

\subsubsection{Observation}

In this study, the perception information comes from three parts: (1) V2I communication-based signalized traffic signal information which includes the current SPaT messages; (2) on-board sensors which include three radar units (left-detecting distance $d_{l}$, right-detecting distance $d_{r}$ and front-detecting distance $d_{f}$) and a front camera (output image size: 320x160 pixels, 50fps); and (3) on-board diagnosis (OBD) port which provides speed $v$ and acceleration value $a$ of the ego-vehicle. 
    
To enhance the learning performance with multiple inputs, we find an efficient way to reduce the complexity of the input observation without losing too much information. For the radar data, we define three variables: the forward warning $w_{f}$, the left warning $w_{l}$ and the right warning $w_{r}$. The definitions of these variables are shown as follows: 
    
\begin{equation}
W_{d}(t) = \frac{ 3+(v(t)-v_{f}(t))^{2}}{2a(t)}
\end{equation}

\begin{equation}
w_{f}(t)=\left\{
\begin{array}{rcl}
0, &{ d_{f}(t)>W_{d}}\\
1, & { d_{f}(t)<=W_{d} }
\end{array} \right.
\end{equation}

\begin{equation}
w_{l}(t)=\left\{
\begin{array}{rcl}
0, &{ d_{l}(t)>W_{l}}\\
1, & { d_{l}(t)<=W_{l} }
\end{array} \right.
\end{equation}

\begin{equation}
w_{r}(t)=\left\{
\begin{array}{rcl}
0, &{ d_{r}(t)>W_{r}}\\
1, & { d_{r}(t)<=W_{r} }
\end{array} \right.
\end{equation}
where $W_{d}, v_{f}, W_{l},W_{r}$ represent the forward warning threshold distance, forward vehicle velocity, left-warning threshold distance and right-warning threshold distance separately. In the simulation, $W_{l},W_{r}$ are both set as 2m. Thus, the observation space $\mathbf{O}$ is composed of a 12-dimensional vector:
\begin{equation}
\mathbf{O} = \{t_{g}, t_{y}, t_{r}, w_{f}, w_{l},w_{r}, w_{c}, d_{r}, d_{f}, v_{f}, v, a \}  
\end{equation}
where the description of the elements in $\mathbf{O}$ is shown as Table~\ref{tab:observation space}.
\begin{table}[]
    \centering
    \caption{The description of the observation space $\mathbf{O}$.}
    \label{tab:observation space}
    \begin{tabular}{c|c}
    \hline
         Element & Description \\\hline
         $t_{g}$ & duration time of green phase \\
         $t_{y}$ & duration time of yellow phase \\
         $t_{r}$ & duration time of red phase \\
         $w_{f}$ & dangerous warning for the forward side \\
         $w_{l}$ & dangerous warning for the left side \\
         $w_{r}$ & dangerous warning for the right side \\
         $w_{c}$ & warning when collision happens \\
         $d_{r}$ & the remain distance to the stop line of the intersection \\
         $d_{f}$ & the distance to the forward vehicle in the current lane\\
         $v_{f}$ & the velocity of the forward vehicle in the current lane \\
         $v$ & the velocity of ego-vehicle \\
         $a$ & the acceleration of ego-vehicle\\\hline
    \end{tabular}
\end{table}

\subsubsection{Energy Model}

For the energy consumption model (ECM), we applied a previous developed and calibrated model deriving from on-road  driving  data of a 2013 NISSAN LEAF \cite{ye2016hybrid}. The original energy consumption model is shown as:
        
\begin{equation}
\label{Dueling}
E(t) = E(v(t), a(t), \alpha)
\end{equation}
where $v(t), a(t), \alpha$ represent the instant speed ($m/s$), acceleration ($m/s^{2}$), and road grade ($rad$) separately. In this study, the road grade is set to zero. Additionally, in the original model, braking-energy regeneration is included. In this paper, however, we redefined the original model which is shown below:
\begin{equation}
E_{energy}(t)=\left\{
\begin{array}{rcl}
E(t), &{ a(t) \geq 0}\\
0, & { a(t)<0 }
\end{array} \right.
\end{equation}
where $E_{energy}$ represents the energy consumption model applied in our framework. The main reason to truncate the braking-energy regeneration in this work is that this component will lead the agent driving slowly and finally stop without moving along the training iterations. There may have a better way to consider the braking-energy regeneration in a learning model, but so far in our work, we simply get rid of this part to make the whole learning system running successfully.

\subsubsection{Action}
As the main purpose of this study is to learn an optimal strategy in both longitudinal and lateral driving maneuvers. The output actions are defined as follows. 

For longitudinal maneuvers, the action space $\mathbf{A}_{lon}$ is 

\begin{equation}
\begin{split}
  \mathbf{A}_{lon} = \{ & 1.0a, 0.8a, 0.6a, 0.4a, 0.2a, 0.0, \\     
  & -0.2a, -0.4a, -0.6a, -0.8a, -1.0a \}  
\end{split}
\end{equation}
where $a$ represents the maximum acceleration. 

For the lateral maneuver, the target lane action space is 
\begin{equation}
    \mathbf{A}_{lat} = \{ -1, 0, 1 \}
\end{equation}
where $-1, 0, 1$ represent the left lane, the current lane, and the right lane respectively.

To enhance the learning performance, we reduce the dimension of the total action space $\mathbf{A}$ by fixing speed during lane-changing maneuvers (i.e. the ego-vehicle can only change lanes when the longitudinal acceleration is zero). In this way, the action space will be reduced from a 33-dimensional vector to a 13-dimensional vector.

\section{HRL-based Eco-Driving Framework}
\subsection{System Architecture}

\begin{figure}[ht]
    \centering
    \includegraphics[width = 0.47\textwidth]{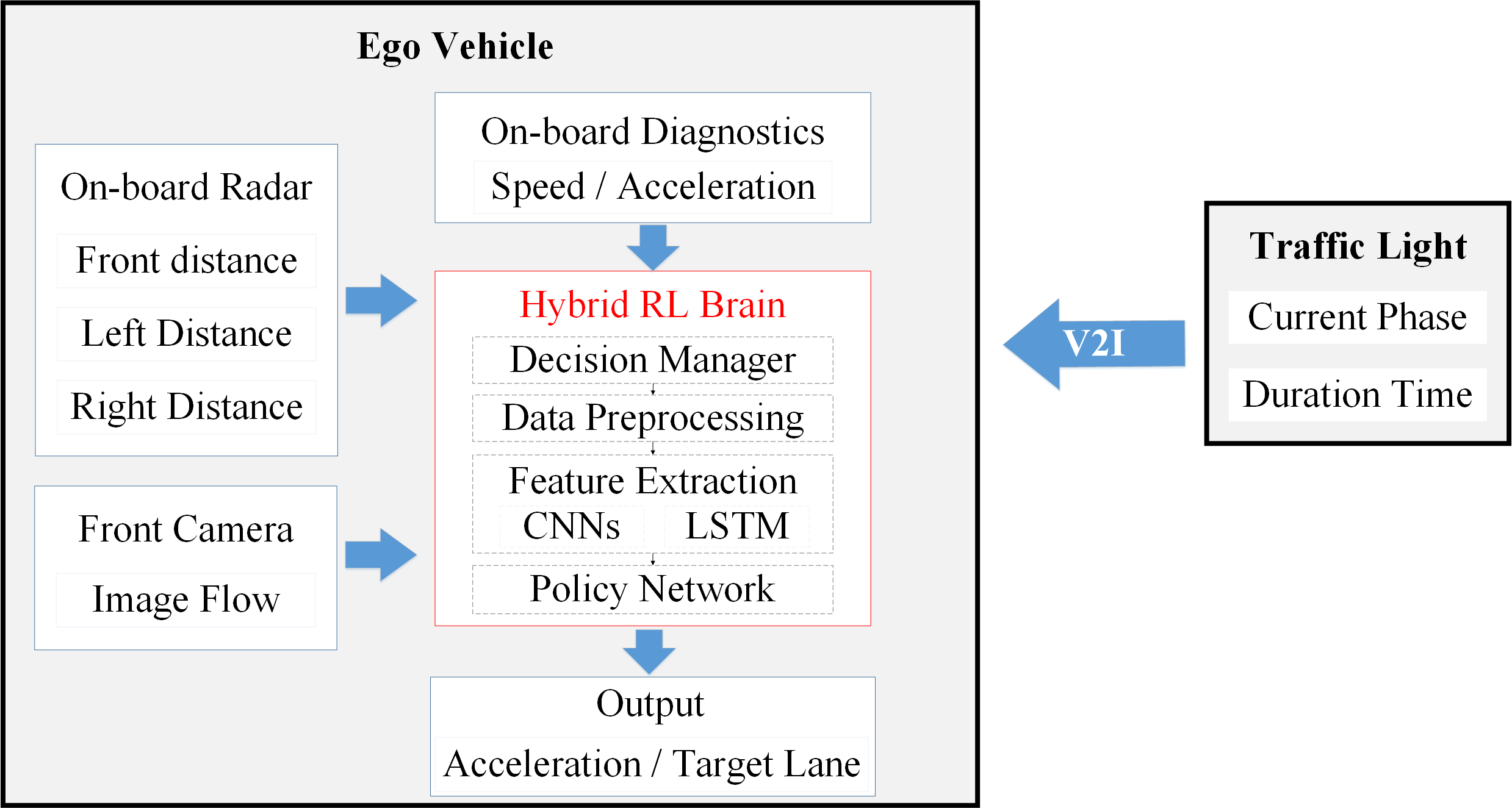}
    \caption{The HRL-based Eco-Driving system architecture.}
    \label{fig:system}
\end{figure}

In this research, we proposed the HRL-based CAV eco-driving framework and algorithms for electric passenger vehicles. Figure~\ref{fig:system} illustrates the key components of the hierarchical-RL based CAV eco-driving system which consists of several components as described briefly below.

\textbf{1. On-board computer:} it houses the hierarchical-RL brain which is the decision-making center of the whole system. The brain will receive and process the data. Then it will return the longitudinal acceleration and target lane information to the vehicle control center.

\textbf{2. On-board radar}: As defined in previous section, three radar units are mounted in the ego-vehicle. One is installed in front of the vehicle to detect the front distance and front-vehicle velocity.  The other two are installed at the left side and right  side of the vehicle to detect the left distance and right distance. The range of the radar units is set to 100 meters. The radar information is sent to the on-board computer as part of the agent observation.

\textbf{3. On-board camera}: this is installed in the front of the vehicle. The direction of the camera is the same as the vehicle's driving direction. As the key source of the traffic perception, the camera information will be sent to the on-board computer. 

\textbf{4. On-board diagnostics (OBD)}: this component can get the instant speed and acceleration information and then the message will be sent to the on-board computer as part of the observation.

\textbf{5. Traffic light and road-side unit}: the signal controller will send the real-time traffic light information to the road-side unit. Then the road-side unit will send the information to any CAV within its communication coverage. 
    
As the most crucial component in the system, the Hierarchical-RL brain aims to drive through an intersection  with  less  time and energy consumption by generating appropriate instant longitudinal acceleration and target lane. As we have discussed above, driving through intersection is actually a logically complex task. There are at least four sub-tasks:
\begin{enumerate}
    \item cruise control that avoids collision with other vehicle; 
    \item lane-changing that improves the driving efficiency in the mixed traffic; 
    \item  Stop-in-red  maneuver that  stops  the  vehicle  before  the stop line when the light is red or yellow; 
    \item Start-in-green mechanism that accelerates the vehicle when the signal turns green. 
\end{enumerate}

So far, it is nearly impossible for one RL algorithm to handle such a multi-phase task. In this research, we proposed a HRL brain which combines the rules and RL algorithm. The Hierarchical-RL brain consists of several components which are illustrated as following sections.

\subsection{HRL Framework}

\begin{figure}[!ht]
    \centering
    \includegraphics[width = 0.48\textwidth]{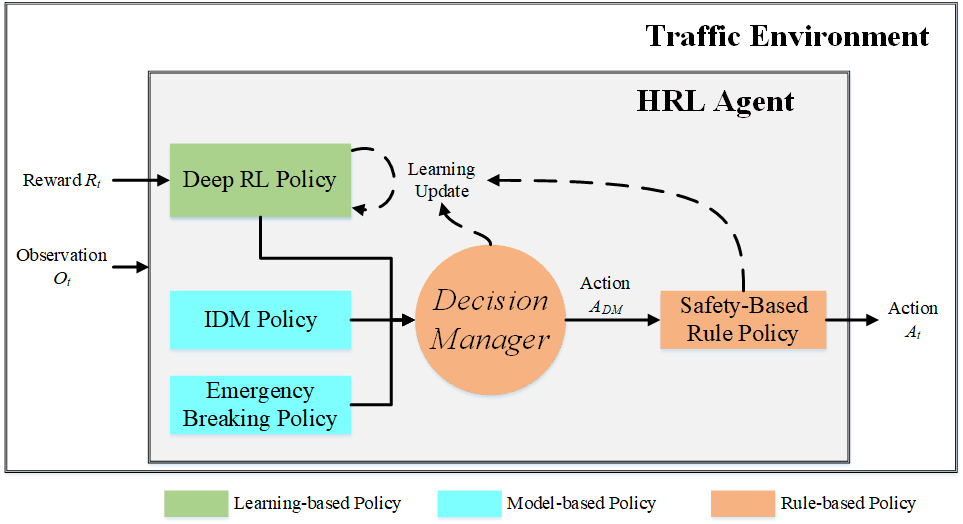}
    \caption{The visualization for the HRL framework.}
    \label{fig:HRL}
\end{figure}

The framework of the HRL method is shown as Figure~\ref{fig:HRL}. The HRL Agent interact with traffic environment through reward $R_{t}$, observation $O_{t}$ and action $A_{t}$. In the HRL agent, there are five main components, which are (1) Deep RL Policy: generating longitudinal and lateral driving actions; (2) IDM Policy: generating longitudinal actions according to the IDM model; (3) Emergency Breaking (EB) Policy: generating longitudinal actions according to the EB Policy; (4) Decision Manager: combining the first three driving policies according to current 
driving situation; and (5) Safety-Based Rule Policy: final safety-based constraints for generating action $A_{t}$.

The IDM policy is discussed in section II, the Decision Manager, Deep RL Policy, EM Policy and Safety-Based Rule Policy will be discussed in the following sections.

\subsection{Data Preprocessing}

\begin{figure}[!h]
    \centering
    \subfigure[Image data preprocessing.]{
    \includegraphics[width=0.45\textwidth]{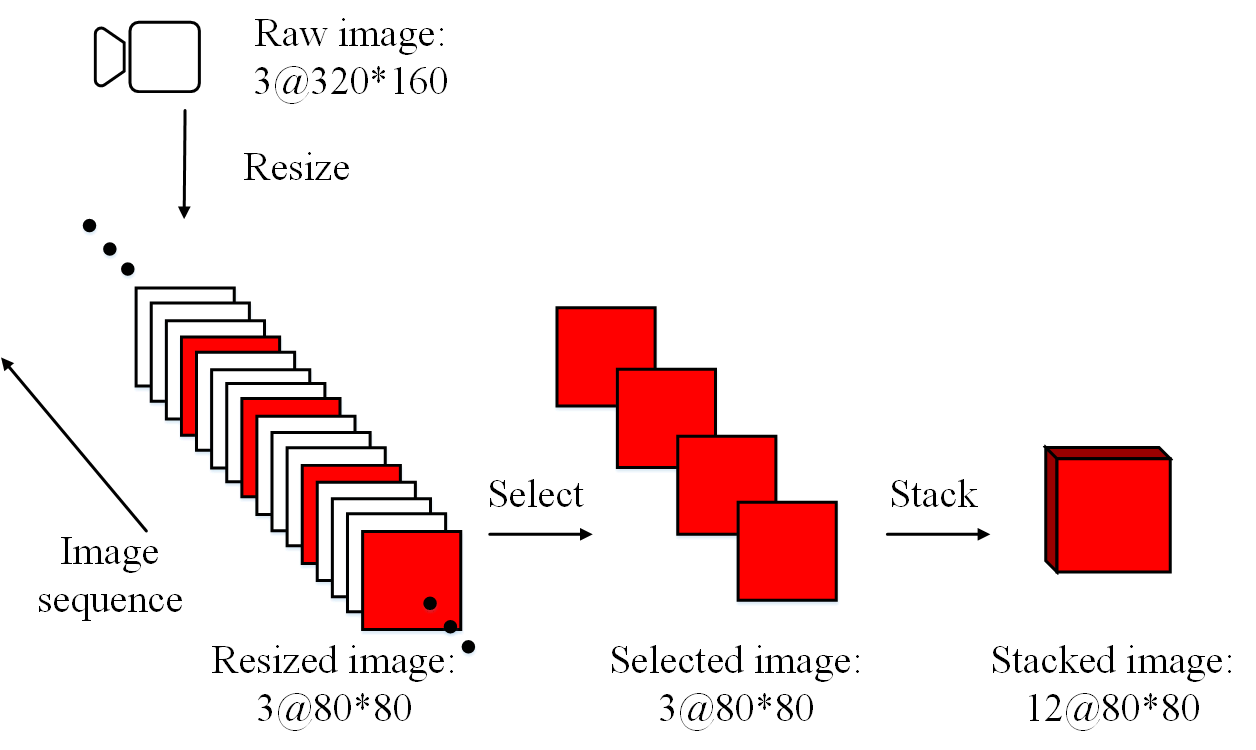}
    \label{fig:image preprocess-a}
    }
    \subfigure[Logical data preprocessing.]{
    \includegraphics[width=0.45\textwidth]{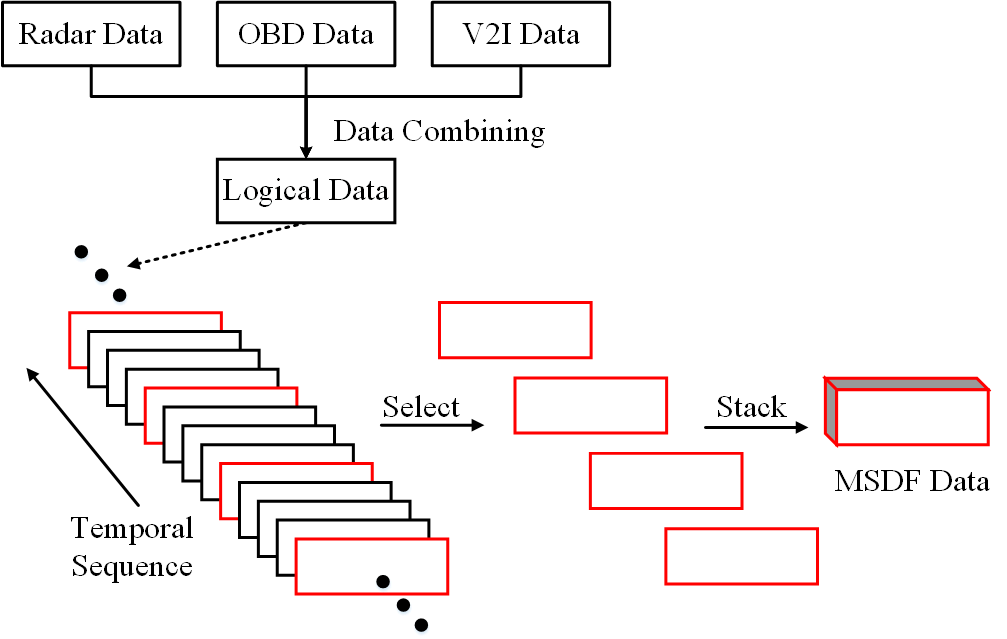}
    \label{fig:image preprocess-b}
    }
    \caption{The select-stack process for data preprocessing.}
    \label{fig:image preprocess}
\end{figure}

In this research, the observation input of ego-vehicle consists of (1) a 50fps raw image flow and (2) a 13-dimensional vector for logical data. In most cases, the raw image data will not be fed into the training network directly because resizing raw images into a compatible and normalized format for the convolutional neural network can greatly improve the computational speed and learning performance. Besides, the purpose of this study is to explore the ability of the HRL to generate the eco-driving strategy for CAV. Hence, considering the driving strategy relies on both spatial and temporal information, instead of resizing the images, we also need to transform multiple single-frame images into a multi-frame spatiotemporal data format (MSDF). Specifically, in this study, to include more information without making the input data too heavy (having too many frames in one format), we apply a select-stack preprocess method. Figure~\ref{fig:image preprocess} shows more detail about the preprocessing for image data and logical data.

For image data, the preprocess includes four steps as below: 
\begin{enumerate}
    \item  Recording the raw image flow based on the time sequence; 
    \item Resizing every frame of the raw image flow into an $80\times80\times3$ format. 
    \item Selecting one frame out of $N_{select}$ frames; \item Stacking $N_{stack}$ frame of images into a higher dimensional data format: $80\times80\times3 \times N_{stack}$. 
\end{enumerate}

In addition, this select-stack preprocess method is also applied to logical data, which is shown as Figure~\ref{fig:image preprocess-b}. 
Specifically, in this study, the $N_{select}$ and $N_{stack}$ are both set as 4. 

Through the above preprocessing method, the training network could get spatiotemporal observation data, which is of considerable significance on the driving strategy learning procedure.

\subsection{Decision Manager}

\begin{figure*}[!ht]
    \centering
    \includegraphics[width = 0.75\textwidth]{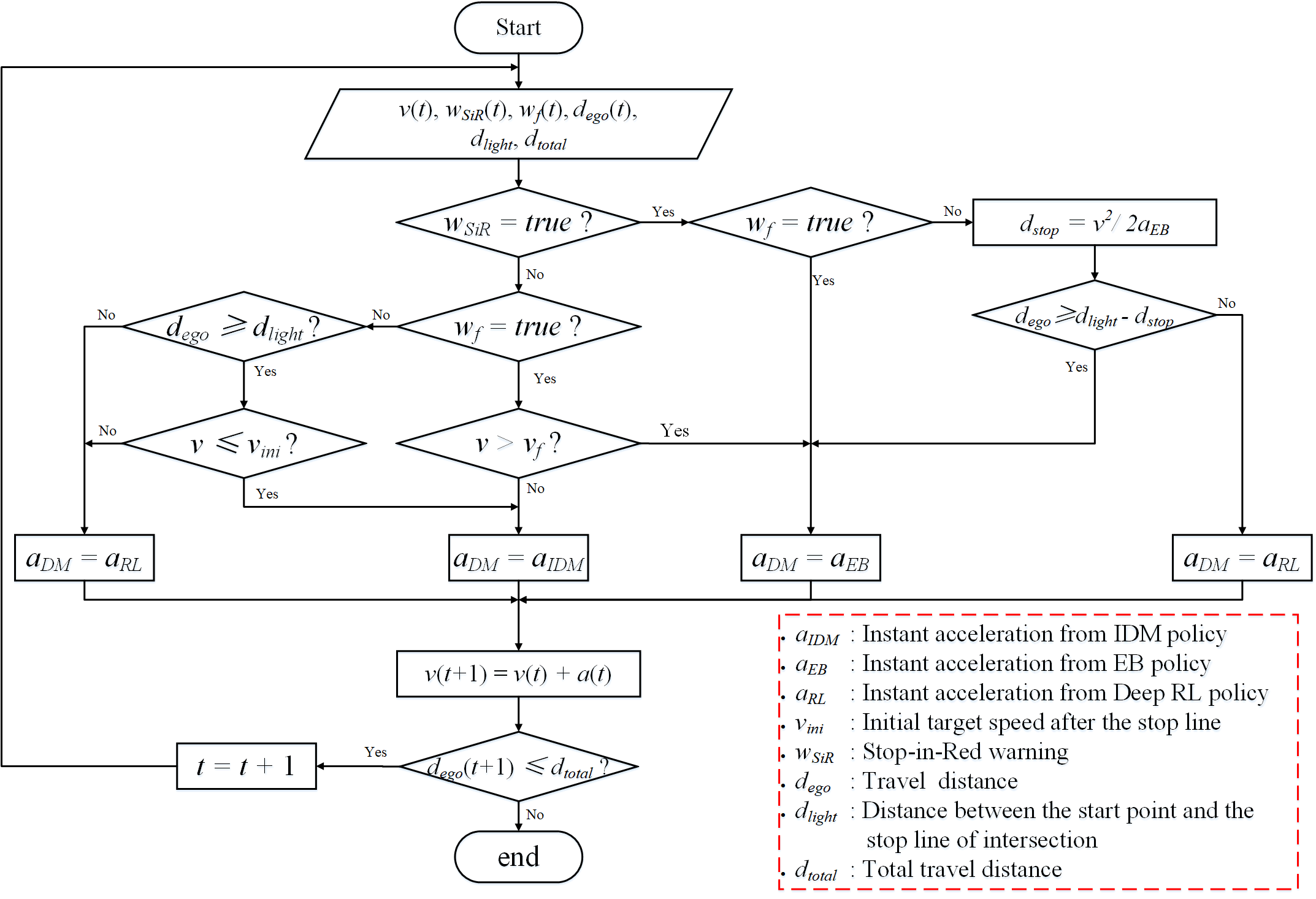}
    \label{fig:manager}
    \caption{The architecture of the Decision Manager.}
    \label{fig:DecisionMnager}
\end{figure*}
    
\begin{figure}[!h]
    \centering
    \includegraphics[width=0.5\textwidth]{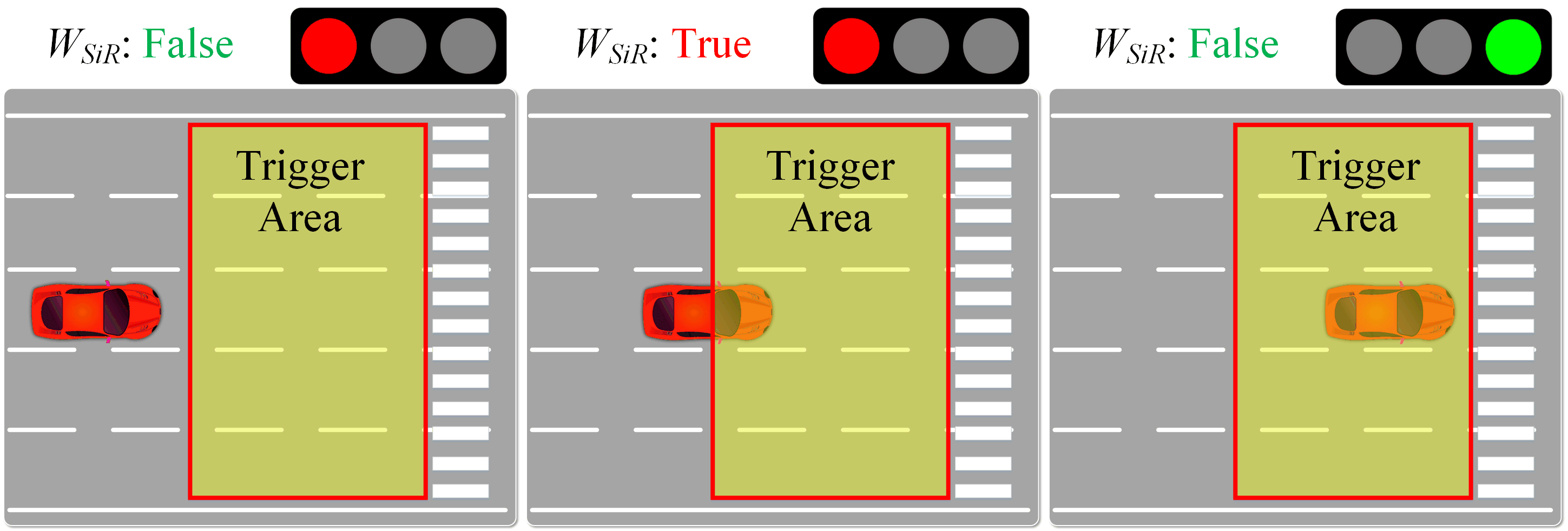}
    \caption{The definition of one key variable in Decision Manager: The Stop-in-Red warning $W_{SiR}$.}
    \label{fig:trigger}
\end{figure}
The key component of the Hierarchical-RL framework is the decision manager which combines the manual-designed rules and the RL algorithm to endow the framework with the ability to handle complex driving tasks. In the decision manager, the driving process is divided into different running situations according to the immediate situations of ego-vehicle and traffic signal. In this paper a warning factor, named Stop-in-Red warning $W_{SiR}$, is designed to differentiate the driving situations. The definition of $W_{SiR}$ and the architecture of the decision manager are shown in Figure~\ref{fig:trigger} and Figure~\ref{fig:DecisionMnager}, respectively.

According to Figure~\ref{fig:trigger}, when the vehicle enters the $W_{SiR}$-trigger area and the phase of current signal is red or yellow, the $W_{SiR}(t)$ is set as True. On the other hand, if the vehicle does not enter the trigger area or the current phase is green, the $W_{SiR}(t)$ is set as False. The main purpose of Stop-in-Red warning is to build a stable interaction between vehicles and traffic signals. As the aforementioned discussions, it is hard for a single RL method to learn complex driving tasks with multiple driving stages. Thus, utilizing the Stop-in-Red warning, the Decision Manager can dynamically combining learning-based and model-based control methods to achieve a complex driving task.

Therefore, the IDM policy and a modified emergency braking policy (EB Policy) are integrated into the decision manager. According to Figure~\ref{fig:DecisionMnager}, the IDM model's output will be activated when the ego-vehicle need to start moving at an intersection when the phase of current traffic signal becomes green. The EB model's output will be activated when the vehicle's front distance is shorter than the EB threshold. In this paper, the EB model is defined as $ \Dot{x}_{lon}(t+\delta t) = \Dot{x}_{lon}(t)+a_{EB}\cdot\delta t$, where $a_{EB}, \delta t$ are set as $-5.0m/s^{2}, 0.02s$.

\subsection{Safety-Based Rule Policy}
In order to make the whole driving strategy safer, we design a Safety-Based Rule (SBR) Policy to finally check the output from safety perspective. The definition of SBR model is shown as Algorithm \ref{alg:SBR}.

\begin{algorithm}
\caption{The definition of SBR model.}
\label{alg:SBR}
\begin{algorithmic}[1] 
    \Require  $a_{DM}(t)$: The action from Decision Manager; \newline
    $a_{EB}$: The action from EB Policy.
     $w_{f}(t)$, $w_{l}(t)$, $w_{r}(t)$: Three near-warning factors from the forward side, left side, and right side;
    \Ensure $a(t)$: The longitudinal and lateral actions.
    
    \Function{SBR}{$a_{DM}(t)$, $a_{EB}$, $w_{l}(t)$, $w_{r}(t)$, $w_{f}(t)$}
        \State $a_{DM}^{lat}(t) \gets $ The lateral component of $a_{DM}$;
        \State $a_{DM}^{lon}(t) \gets $ The longitudinal component of $a_{DM}$
        \If{$w_{f}(t) == True$}
            \State $a_{lon}(t) \gets a_{EB}$;
        \Else
            \State $a_{lon}(t) \gets a_{DM}^{lon}(t)$;
        \EndIf 
        \If{\text{$w_{l}(t)$, or $w_{r}(t)$ is conflicted with $a_{DM}^{lat}(t)$}}
            \State $a_{lat}(t) \gets 0$;
        \Else
            \State $a_{lat}(t) \gets a_{DM}^{lat}(t)$;
        \EndIf 
        \State $a(t) \gets$ Combining $a_{lat}, a_{lon} $;
    
    \algorithmicreturn  \quad$a(t)$;
    \EndFunction
\end{algorithmic}
\end{algorithm}

From Algorithm \ref{alg:SBR}, the final output $a(t)$ is confined by the safety warning factors: $w_{l}(t)$, $w_{r}(t)$, and $w_{f}(t)$. For example, if the $a_{DM}(t)$ means “lane change to the right lane", but the $w_{r}(t)$ is “True", then the action from $a_{DM}(t)$ will not be executed and a relevant penalty will be added to the reward function (more details about this penalty are discussed in next section). Similarly, if the forward warning factor $w_{f}(t)$ is “True", the current longitudinal action will be replaced by emergency braking (EB) model. Hence, utilizing the SBR model, the final action $a(t)$ will always stay in a safe range. 

\subsection{Deep RL for Long-term Eco-Driving}

According to the above discussion, the eco-driving approach of going through an intersection can be formulated as an Markov Decision Process (MDP) in which the agent interacts with the environment. In this research, we applied Dueling Deep Q Network (Dueling DQN) as our basic RL framework and in terms of Long-term Eco-driving tasks, we proposed an innovative reward function named Long-Short Term Reward (LSTR) model to better evaluate the current situation from both long-term factors and short-term factors by considering two conflicting purposes (i.e., saving energy and pursuing speed). The mechanism of Dueling DQN and the LSTR model are discussed below.

Deep Q Network \cite{Mnih2015Human} is a typical deep RL algorithm that uses a deep neural network to predict the value function of each discrete action. Specifically, the input of DQN is observation state $o_{t}$, and the output is the evaluation value $Q(o_{t}, a_{t})$ corresponding to each action $a_{t}$ in space $\mathbf{A}$ under current state $o_{t}$. Then, according to the $\epsilon-greedy$ algorithm, an action is selected from the action space $\mathbf{A}$. After the execution of action $a_{t}$, a reward $r_{t}$ and an observation state $o_{t+1}$ can be get from the environment. In this paper, prioritized experience replay \cite{Schaul2015Prioritized} algorithm is applied to solve the problem of correlation and non-static distribution.

Dueling DQN \cite{wang2015dueling}, which has outperformed DQN in vision related learning tasks, is constructed with two streams which separately estimate (scalar) state-value and the advantages of each action. Dueling DQN is designed to aggregate the states-value $V$ and action advantages $A$, shown in equation \ref{eq:dueling}.
\begin{equation}
\begin{split}
    Q( O_{t}, A_{t}; \theta, \alpha, \beta) & = V(O_{t}; \theta, \beta) + A(O_{t}, A_{t}; \theta, \alpha) \\
     & - \frac{1}{|A|} \sum_{A_{t}}A(O_{t}, A_{t}; \theta, \alpha)
     \label{eq:dueling}
\end{split}
\end{equation}
where $\alpha$ represents the parameters of $A$ (the advantage function). Besides, $\beta$ represents the parameters of $V$ (the state-value function) and $\theta$ is parameters of neural network.


In this research, the most challenging part for the policy learning is to find an optimal point between two conflicting factors: (1) saving energy and (2) pursuing speed under mixed dense traffic conditions. Another challenging part is to learn complex driving strategy with long insight in multi-phases driving tasks. For long insight, it represents that the agent need to evaluate the current situation with very long-term benefit (e.g, whether it can pass the intersection without stop, receiving total energy consumption and travel time, etc.). For multi-phase driving task, it means from the upstream to passing the intersection, the agent needs to adjust to several different operating logic.

\begin{algorithm}
\caption{The descriptions for LSTR model .}
\label{alg:reward}
\begin{algorithmic}[1] 
\Require $v(t)$: The current speed of ego-vehicle; \newline
        $a(t)$: The current acceleration of ego-vehicle; \newline
        $d_{r}(t)$: The remain distance to the stop line; \newline
        $P_{c}(t)$: The current phase of traffic signal; \newline
        $T_{r}(t)$: The remaining time of the current signal phase;
\Ensure  $R(t)$: The reward value at current time step $t$;
\Function{LSTR}{$v(t)$, $a(t)$, $d_{r}(t)$, $P_{c}(t)$, $T_{r}(t)$}
    \State $d_{pridction} \gets v(t)\cdot T_{r}(t)$; 
    \State $R_{velocity}(t) \gets  \dfrac{v(t) - v_{min}}{v_{max} - v_{min}}$;
    \label{ code:fram:velocity}
    \While{$d_{r}(t) > 0$}
        \If{$a(t) > 0$;}
            \State $R_{energy}(t) \gets E(a(t),v(t),\alpha)$;
        \Else
            \State $R_{energy}(t) \gets 0$;
        \EndIf
        \label{ code:fram:energy }
        \State $R_{time}(t) \gets R_{time}(t-1) - \delta t$
        \If{Lane-change action happens;}
            \State $R_{lanechange}(t) \gets -0.1;$
        \EndIf
        \label{code:fram:lanechange}
        
        \If{Dangerous action happens;}
            \State $R_{danger}(t) \gets -0.5;$
        \EndIf
        \label{code:fram:danger}
        \State $ R(t) \gets w_{v}R_{velocity}(t) + w_{e}R_{energy}(t) + w_{t}R_{time}(t) + w_{lc}R_{lanechange}(t)  + w_{d}R_{danger}(t) + w_{GP}R_{GP}(t)$
        \label{code:fram:sum}
     
    \algorithmicreturn \quad$R(t)$; 
    \EndWhile
\EndFunction

\end{algorithmic}
\end{algorithm}

In order to solve aforementioned challenges, this study proposed a long-short term reward (LSTR) model, which not only considers the instance values such as speed, lane-changing actions, and instant energy consumption, but also includes long-term indicators based on predefined rules. The two conflicting factors in this issue are the short-term factors (instant speed, energy consumption) and the long-term factors (total travel time and total energy consumption). We designed some instant reward principles which include indications for long-term benefit, which are shown below. 

\begin{itemize}
\item When the current phase $P_{c}$ is red or yellow and ego-vehicle cannot pass the intersection with its current speed, then it shouldn't accelerate.
\item When the current phase $P_{c}$ is red or yellow and ego-vehicle may pass the intersection with current or higher speed, then try to accelerate.
\item When the current phase $P_{c}$ is green and the ego-vehicle cannot pass the intersection with its current speed, then it shouldn't accelerate.
\item When the current phase $P_{c}$ is green and the ego-vehicle may pass the intersection with current or higher speed, then try to accelerate.

\end{itemize}

\begin{algorithm}
\caption{The definition of $R_{GP}$.}
\label{alg:lightReward}
\begin{algorithmic}[1] 
    \Require $v(t)$, $a(t)$, $d_{r}(t)$, $T_{r}(t)$, $P_{c}(t)$
    \Ensure The Green-Pass Reward $R_{GP}(t)$
    
    \Function{$R_{GP}$}{$v(t)$, $a(t)$, $d_{r}(t)$, $T_{r}(t)$, $P_{c}(t)$}
        \State $d_{prediction}\gets v(t) \cdot T_{r}(t)$
        \While{$d_{r}(t) > 0$}
            \If{Current signal phase $P_{c}(t)$ is GREEN}
                \If{$d_{prediction(t)} > d_{r}(t)$} 
                    \State Case $a(t) > 0: R_{GP}(t)\gets 0.2$
                    \State Case $a(t) = 0: R_{GP}(t)\gets 0.5$
                    \State Case $a(t) < 0: R_{GP}(t)\gets -1.0$
                \EndIf
                \If{$d_{prediction(t)} > d_{r}(t)$} 
                    \State Case $a(t) > 0: R_{GP}(t)\gets 1.0$
                    \State Case $a(t) = 0: R_{GP}(t)\gets 0.5$
                    \State Case $a(t) < 0: R_{GP}(t)\gets -1.0$
                \EndIf
            \Else
                \If{$d_{prediction(t)} > d_{r}(t)$} 
                    \State Case $a(t) > 0: R_{GP}(t)\gets -1.0$
                    \State Case $a(t) = 0: R_{GP}(t)\gets 0.5$
                    \State Case $a(t) < 0: R_{GP}(t)\gets 0.2$
                \EndIf
                \If{$d_{prediction(t)} > d_{r}(t)$} 
                    \State Case $a(t) > 0: R_{GP}(t)\gets 0.5$
                    \State Case $a(t) = 0: R_{GP}(t)\gets 0.5$
                    \State Case $a(t) < 0: R_{GP}(t)\gets -0.5$
                \EndIf
            \EndIf 

        \algorithmicreturn \quad$R_{GP}(t)$;
        \EndWhile
    \EndFunction
\end{algorithmic}
\end{algorithm}

Basing on these principles, the LSTR model is designed as Algorithm \ref{alg:reward} in which the definition of $R_{GP}$ is further explained by Algorithm \ref{alg:lightReward}. In a nutshell, the $R_{GP}$ reflects the long-term benefit for reaching the stop line of the intersection when the light is green. 

\begin{table}[!ht]
\caption{The values of coefficients in Algorithm \ref{alg:reward}.}
\label{tab:coefficient}
\centering
\begin{tabular}{c|c|c|c|c|c|c}
\hline
     Coefficient & $w_{v}$ & $w_{e}$&$w_{t}$ &$w_{lc}$ & $w_{d}$ & $w_{GP}$  \\ \hline
    Value& 2.0 & 0.1 & 0.002 &3.0& 1.0 &0.1\\
\hline
\end{tabular}
\end{table}

In the LSTR model, $R_{velocity}(t)$, $R_{energy}(t)$, $R_{lanechange}(t)$, and $R_{danger}(t)$ are short-term benefit indicators, while the $R_{time}(t)$, $R_{GP}(t)$ are long-term benefit indicators.

\subsection{Neural Network Structure}

\begin{figure*}[ht]
    \centering
    \includegraphics[width = 0.9\textwidth]{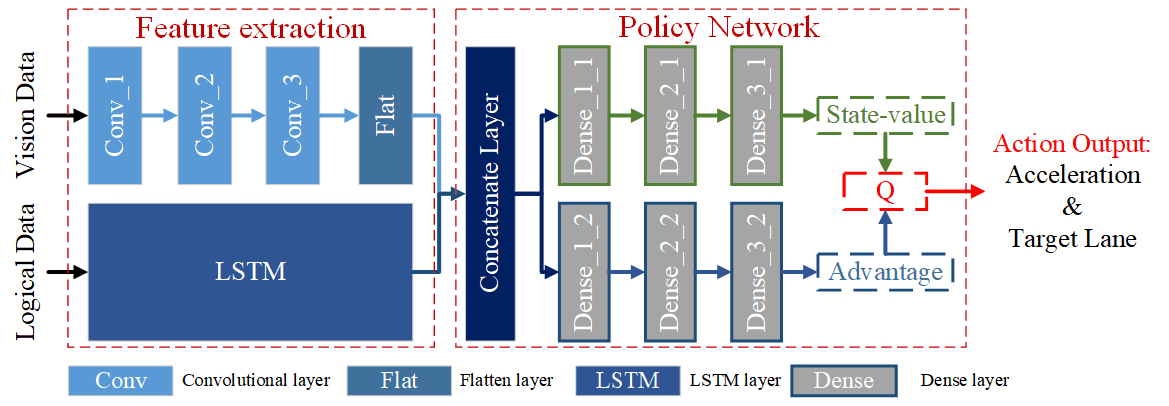}
    \caption{The architecture of the neural network.}
    \label{fig:mainNet}
\end{figure*}

The main purpose of the deep neural network is mapping the current observation state O to the best action value A in action space. Thus, the main network can be divided into two components: (1) the hidden feature extraction network and (2) the policy network which applies Dueling DQN. Figure~\ref{fig:mainNet} illustrates the architecture of the proposed neural network for the applied Dueling DQN model.
       
For the hidden feature extraction network, its main purpose is to extract hidden features from the preprocessed high-dimensional spatiotemporal data. As discussed above, the network input consists of (1) image flow and (2) high dimensional vector. In order to extract features from these different type of data, we designed a multi-channel feature extraction network with different kinds of parallel neural networks. 

For the image flow input, a convolutional neural network (CNN) \cite{albawi2017understanding} stream is designed to extract the vision-based input data. On the other hand, the radar, OBD, and V2I information (named as logical data) are integrated into a 13-dimension vector. After the selected-stack preprocess, the logical data  are transformed into a time-series based temporal data. Long-short term memory (LSTM) network \cite{hochreiter1997long} is then applied for its capability in dealing with temporal data series. The features from both CNN and LSTM streams are combined through a concatenate layer and then a dense-layer-based policy network is designed by applying Dueling DQN. Figure~\ref{fig:mainNet} illustrates the two streams of the dense layer. The upper stream is used to extract state value, while the output of lower stream represents the advantage of current state. Then, the Q-table representing the future gain for each possible action $a_{i} \in \mathbf{A}$ under current state $o_{i}$ is generated based on the state-value and advantage. Finally, the action with maximum Q-value will be selected as output action at this time step. Table \ref{table:netparameters} shows the details of the configuration of the neural network.




\begin{table}[htb]
\centering
\caption{Parameters for Deep Neural Network.}
\label{table:netparameters}
\begin{tabular}{c|c|c|c|c|c}
\hline
\textbf{Layer}&	\textbf{Actuation}&\textbf{Patch size}	&\textbf{Stride}& \textbf{Filter} & \textbf{Unit}		\\\hline
Conv\_1&	ReLU	&	$8\times8$			 & 4 	 & 32 & -		\\
Conv\_2&	ReLU	&	$4\times4$			 & 2 	 & 64 &-		\\
Conv\_3&	ReLU	&	$3\times3$			& 1	 	& 64 & -		\\\hline
LSTM&	-	&	- 	     	 & - 	 & -    & 1024		\\\hline
Dense\_1\_1/2&	ReLU	&	- 	     	 & - 	 & -    & 1024		\\
Dense\_2\_1/2&	ReLU	&	- 			 & - 	 & -    & 256		\\
Dense\_3\_1/2&	ReLU	&	- 			 & - 	 & -    & 128		\\\hline
\end{tabular}
\end{table}

\subsection{Network Update and Hyperparameters}

There are four steps of the network updating process, which are:
\begin{enumerate}
    \item Considering the current state as $o_{t}$ and predicting the $Q_{t}$ value of different actions through the evaluation network.
    \item Choosing the action $a(t)$ with the largest $Q$ value by utilizing $\epsilon-greedy$ policy. 
    \item Generating the $Q(t+1)$ values at time through the target network.
    \item Calculating the loss function and then updating the evaluation network.
\end{enumerate}
    
In addition, at each learning step, the weight coefficients of the proposed network are updated using the adaptive learning rate trick Adam \cite{Kingma2014Adam} in order to minimize the loss function. For the adopted hyperparameters, the learning rate $\alpha$, discount factor $\gamma$, batch size, steps used for observation, replay memory size, steps for target network update, training steps, and test steps are set as 0.00025, 0.99, 64, 10,000, 50,000, 10,000, 2,000,000, 10,000, respectively.

\section{Game Engine Based Simulation Experiments}
\subsection{Unity Engine Based Simulator Design and Development}
 
In order to train the proposed model and evaluate its performance, a simulation platform has been developed in this paper based on Unity Engine and Unity Machine Learning Agents Toolkit (Unity ML-Agents) \cite{juliani2018unity}. Unity is a real-time 3D Game Engine that can generate high-resolution sensor data, like camera, radar, etc. Unity ML-Agents is an open-source platform that enables the simulations to serve as environments for ML training tasks. In this paper, we applied the Unity Engine to build a signalized intersection traffic environment based on previous 
work \cite{bai2019deep, min2018deep}. 

\begin{figure}[!h]
    \centering
    \includegraphics[width=0.37\textwidth]{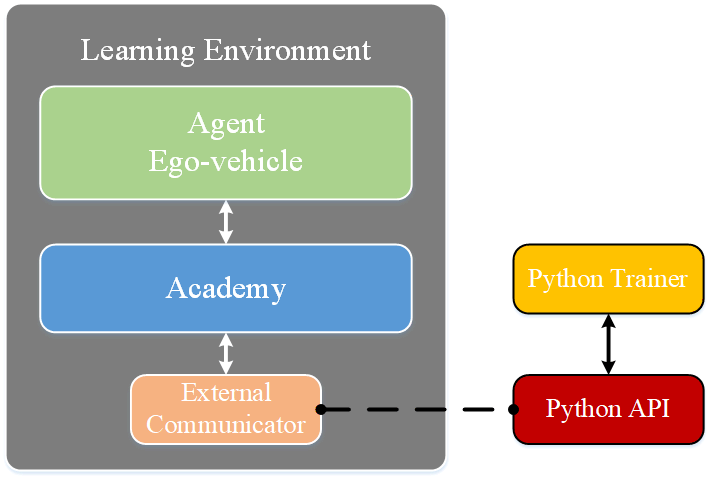}
    \caption{The architecture of the learning environment based on the Unity ML-Agents.}
    \label{fig:unity}
\end{figure}

Unity can provide a virtual reality environment in which the simulated camera, radar, and traffic lights can be developed accordingly. In addition, Unity ML-Agents provides a machine learning toolkit thus we developed a external script named RL brain to construct the deep RL based on Tensorflow \cite{abadi2016tensorflow} by Python. Specifically, the structure of the developed simulation platform based on Unity is shown as Figure~\ref{fig:unity} where the Agent (ego-vehicle) is responsible for collecting observations and executing actions and the Academy is responsible for global coordination of the environment simulation.

\begin{figure}[!h]
    \centering
    \includegraphics[width=0.47\textwidth]{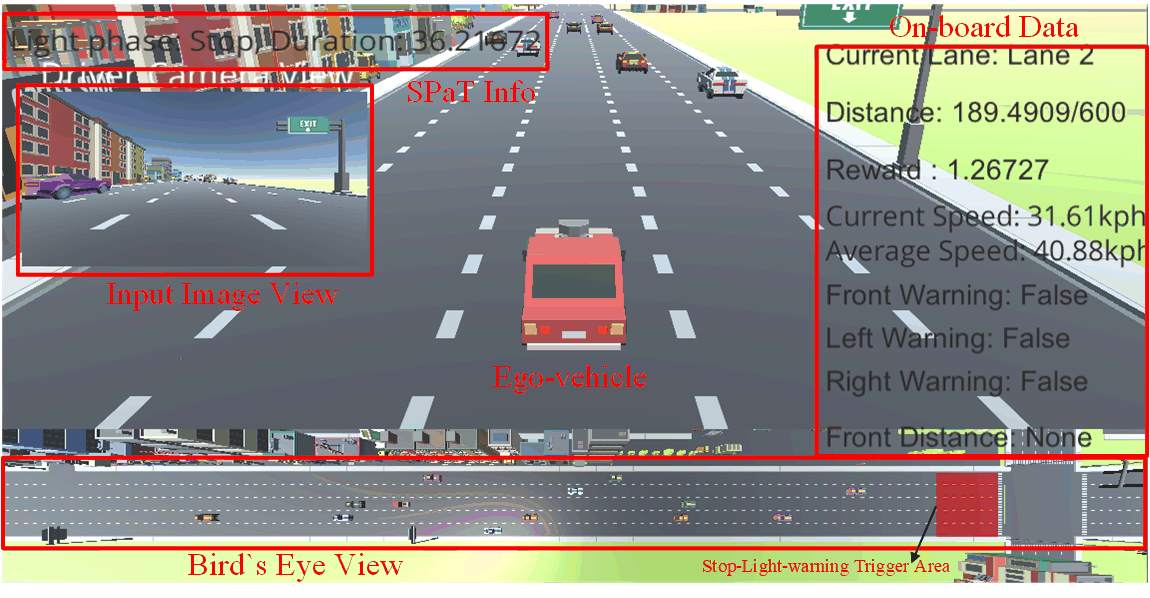}
    \caption{The description of the developed simulator.}
    \label{fig:simulatorView}
\end{figure}

As discussed in the previous section, the scenario in this study is built as a one-direction intersection with 5 lanes and 5 types of vehicles as the traffic. The length of the research area is 550 meters, including a 510-meters upstream section and a 40-meters downstream section. The maximum speed limit is set to 50 kilometers per hour (kph). Figure~\ref{fig:simulatorView} illustrates the main window of the simulator.

\subsection{Model Training and Numerical Testing Design}
     
For the training procedure, the $\epsilon-greedy$  policy is implemented as the exploration policy. The $\epsilon$  is decreased linearly from 1 to 0.00001 over 2 million steps. When the training starts, the initial phase and time of the traffic signal will be randomly selected, which aims to avoid overfitting of the simulation. 
In addition, the simulator is equipped with Intel(R) Core(TM) i7-7700k CPU @ 4.20GHz, 64 GB RAM, and an NVIDIA GTX 1080 GPU. The total training time is around 36 hours.

For the test procedure, we applied IDM method as the baseline and a state-of-art model-based Eco-Driving method, Graph-based Trajectory Planning Algorithm (GBTPA)  to compare the Eco-Driving performance with our HRL method. GBTPA is formulated in \cite{hao2018connected} to find the shortest path or minimum cost path using a graph model in which nodes represent feasible system states and edges define the direct transition from one state to another and the edge weight is the energy cost associated with the transition. These methods are summarized as Table \ref{tab: comparing methods}. Particularly, the rule-based lateral control method is defined as a rule-based lane-changing method, i.e., a lane-change decision will be made if the front distance is shorter than 5m and the target speed is higher than the speed of the front vehicle and the accumulative time is over 3s.

\begin{table}[!h]
\centering
\caption{Summary of compared methods.}
\label{tab: comparing methods}
\begin{tabular}{llll}
\hline
\textbf{Methods} & \textbf{Type}   & \textbf{Lon. Control} & \textbf{Lat. Control} \\ \hline
IDM              & Model based     & IDM                           & Rule                   \\
Graph            & Model based     & GBTPA                         & Rule                   \\
HRL              & Hybrid RL based & HRL                           & HRL                    \\ \hline
\end{tabular}
\end{table}

The simulation scenarios consider different combinations of entry time and initial speed. The entry time varies from  0th to 50th second of the cycle with 10 seconds as the increment (marked as C0, C10, C20, C30, C40, C50). The initial speed varies from 10 kph to 50 kph with 10 kph as the increment (S10, S20, S30, S40, S50). The training and numerical test results are discussed in the next section.

\section{RESULTS}
\subsection{Training Results}


\begin{figure}[!ht]
    \centering
    \subfigure[Average speed per journey.]{
        \includegraphics[width=0.4\textwidth]{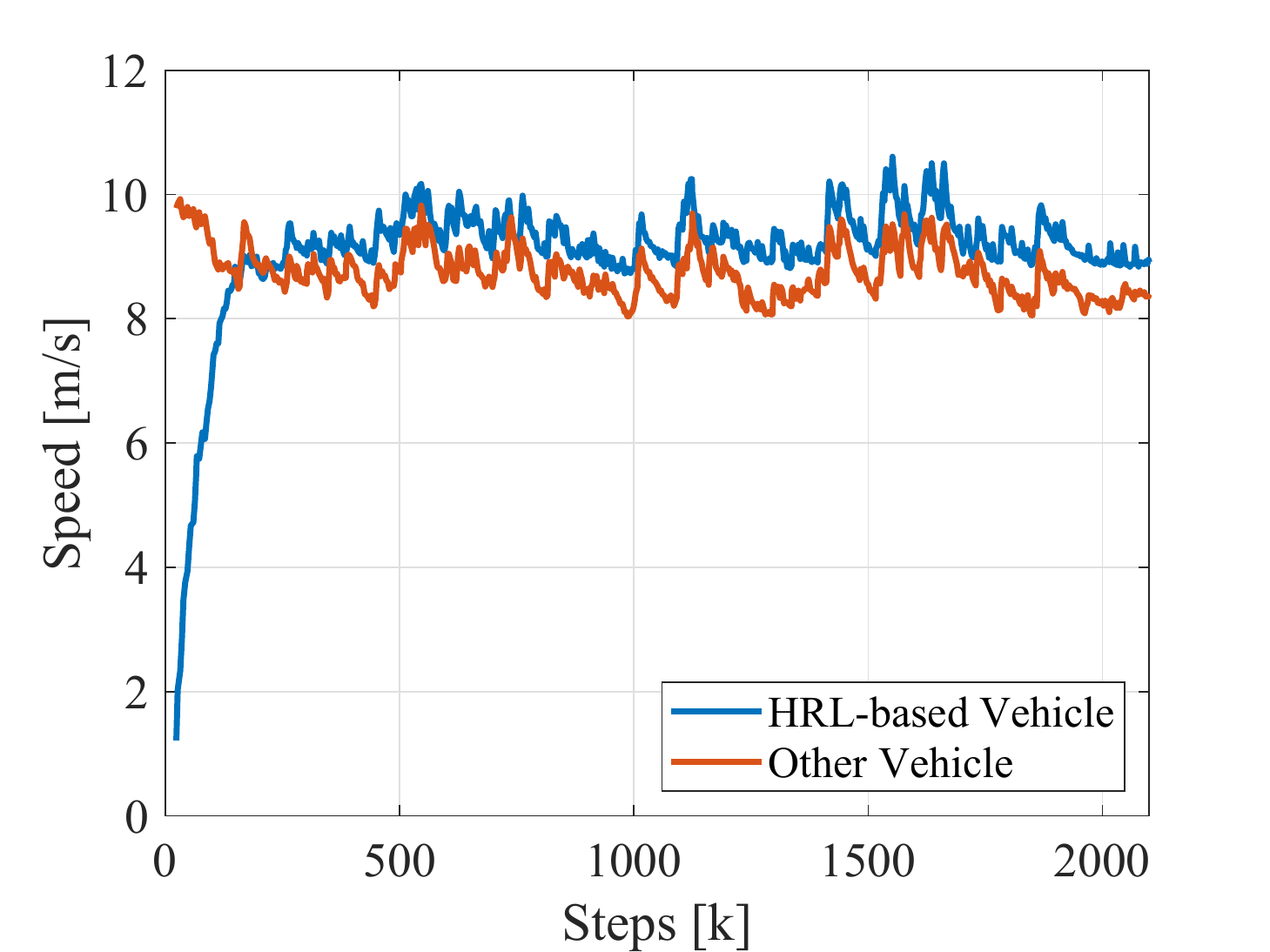}
        \label{fig:training-speed}
    }
    \subfigure[Average energy consumption and $R_{GP}$ per journey.]{
        \includegraphics[width=0.4\textwidth]{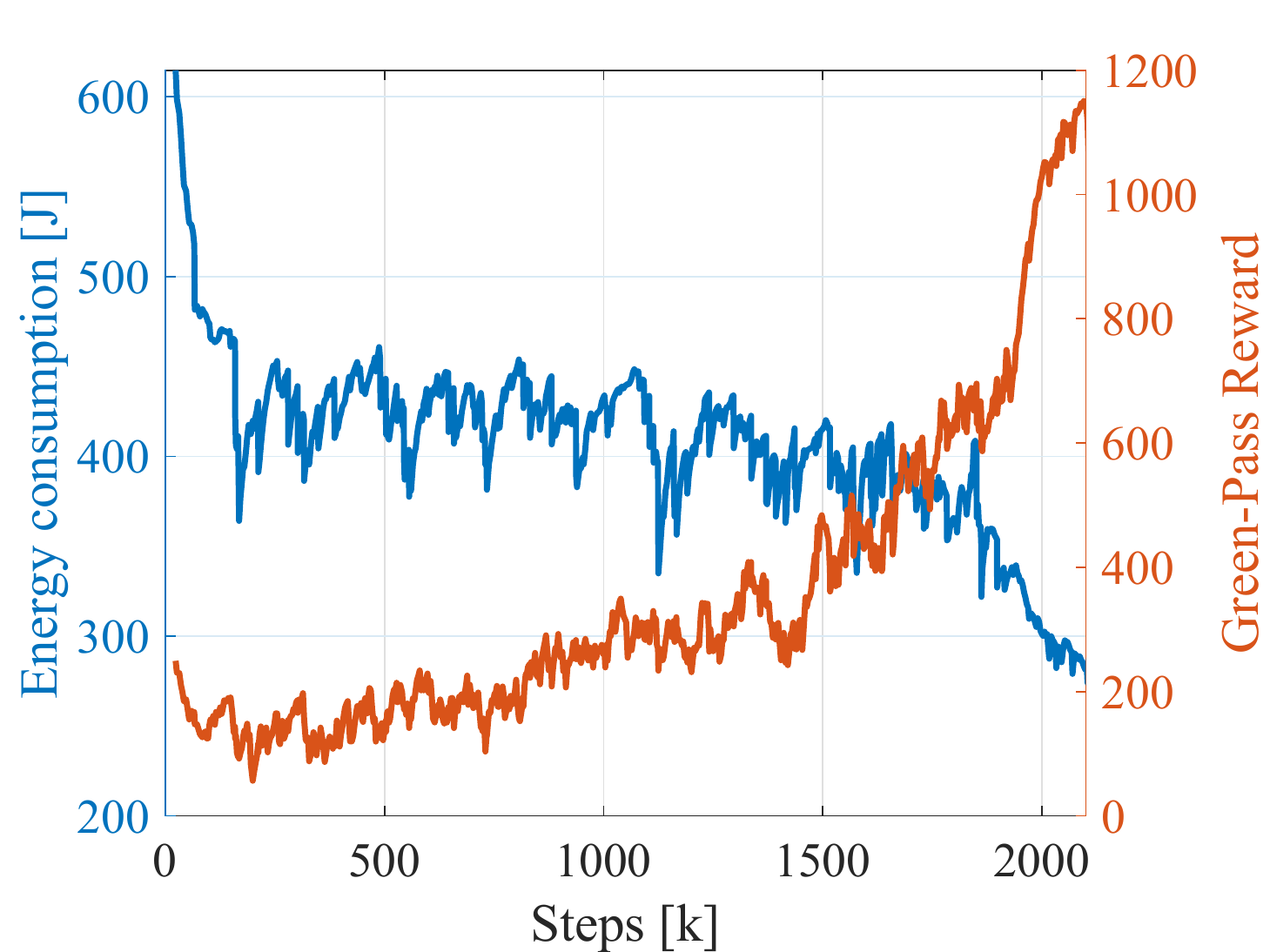}
        \label{fig:training-energy}
    }
    \subfigure[Number of lane-changing actions per journey.]{
        \includegraphics[width=0.4\textwidth]{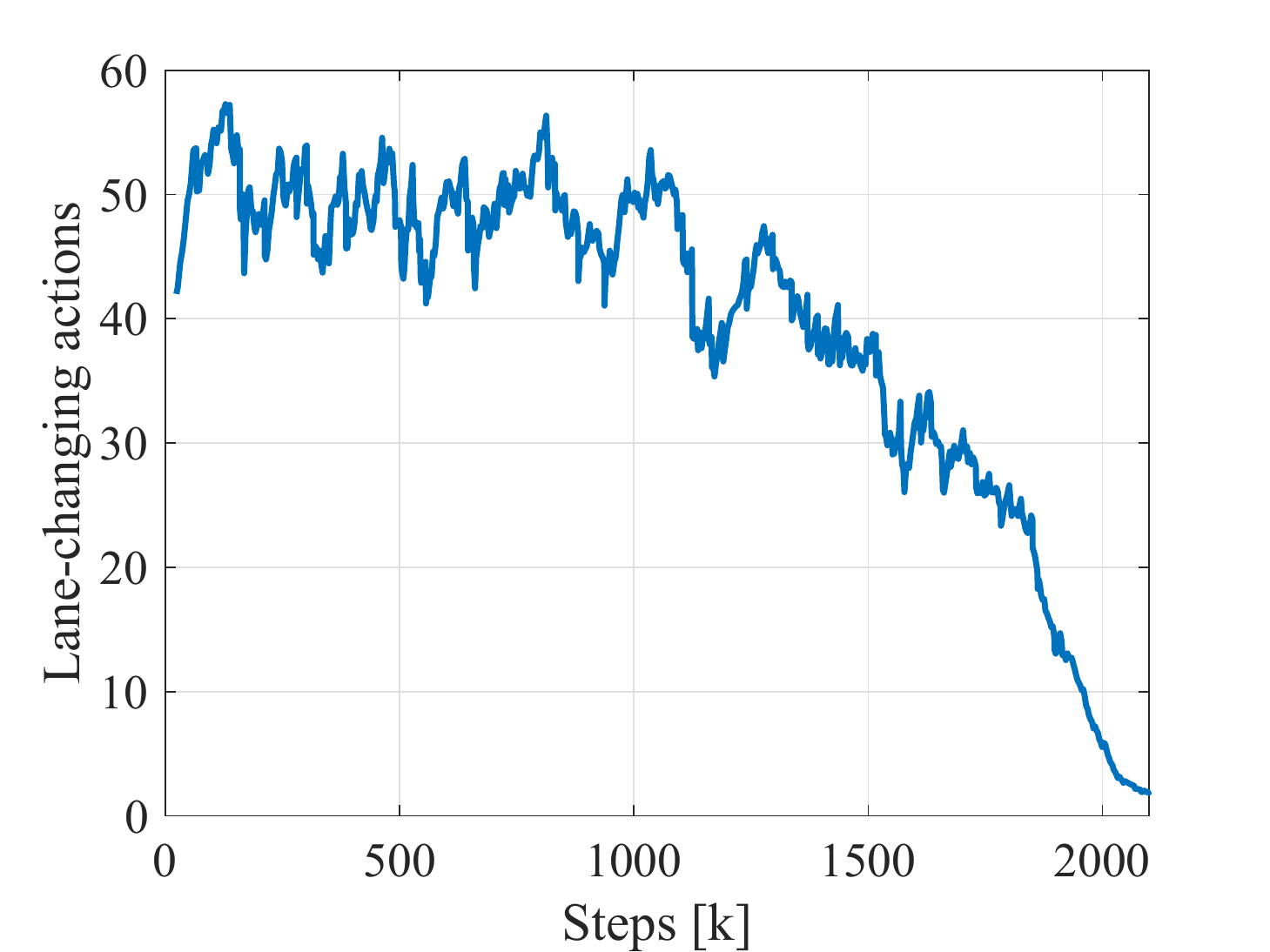}
        \label{fig:training-lc}
    }
    \caption{The training results for (a) average speed per journey, (b) average energy consumption and $R_{GP}$ per journey, and (c) lane-changing actions per journey, respectively.}
    \label{fig:training_results}
\end{figure}

\begin{figure*}[!ht]
    \centering
    \subfigure[Traffic scenario: initial signal time C30 and initial speed 20 kph.]{
        \includegraphics[width=\textwidth]{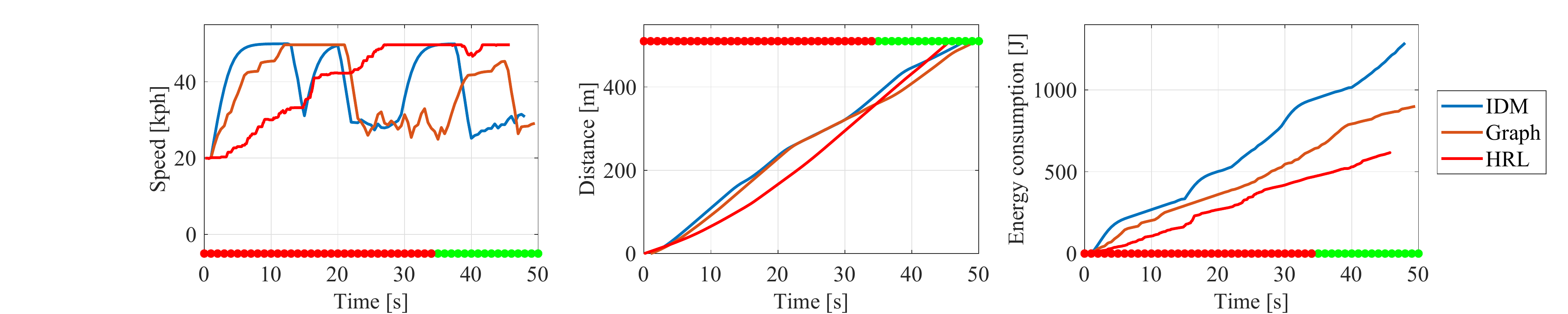}
        \label{fig:C30S20}
    }
    \subfigure[Traffic scenario: initial signal time C40 and initial speed 20 kph.]{
        \includegraphics[width=\textwidth]{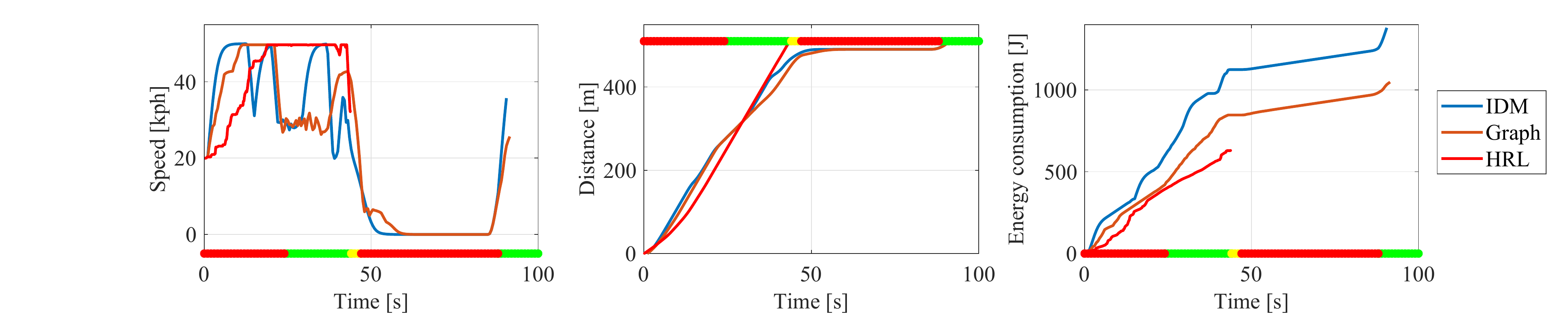}
        \label{fig:C40S20}
    }
    \subfigure[Traffic scenario: initial signal time C50 and initial speed 20 kph.]{
        \includegraphics[width=\textwidth]{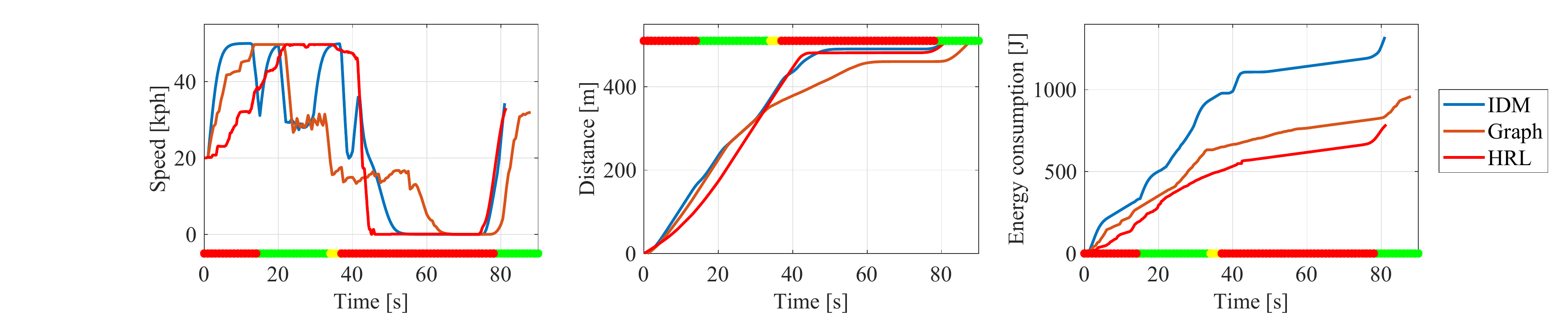}
        \label{fig:C50S20}
    }
    \caption{Spatial-temporal diagrams of the ego vehicle under different traffic scenarios with IDM, Graph and HRL methods, respectively.}
    \label{fig:testing_results}
    
\end{figure*}

Figure~\ref{fig:training_results} illustrates the training  results of HRL method, including the average speed, energy consumption, Green-Pass Reward, and number of lane-changing maneuvers per journey. First, from Figure~\ref{fig:training-speed}, the speed of HRL-based ego-vehicle has increased significantly along the training process and achieved a higher value than the speed of other vehicles. Second, Figure~\ref{fig:training-energy} shows that the energy consumption of HRL method is decreasing via the iteration of training. Specifically, the the curve of Green-Pass reward $R_{GP}$ can further explains why the HRL method can successfully learn an efficient eco-driving policy in such a dilemma, i.e., pursuing both energy-saving and time-efficient driving strategy. After the approximately half way of training steps, the Green-Pass Reward is increasing significantly, as the ego-vehicle is successfully learning about how to interact with the traffic signal intelligently (i.e. balancing the speed and energy consumption). On the other hand, from energy perspective, pursuing a speed-first driving strategy will more likely to cause a longer time for waiting at intersections, which indicates a negative effect on the cooperation between vehicles and intersections. Finally, Figure~\ref{fig:training-lc} shows that the HRL method can also learn how to drive with less lane-changing behaviors reaching a smooth driving strategy.

\subsection{Numerical Testing Results}

Figure~\ref{fig:testing_results} illustrates the spatial-temporal diagrams for comparison testing results under several different traffic scenarios. Particularly, SPaT information is illustrated by green, yellow and red dots to better demonstrate the performance of methods. Additionally, Table \ref{tab:testing results} shows the results of comprehensive assessments for different methods under all different scenarios. According to these testing results, there are three points to discuss:

\textbf{(1) The Acceleration}: the instant acceleration value generated by HRL method can intelligently react to different traffic situations. If high speed will make the vehicle stop at intersections, the HRL will generate a lower acceleration value to have a better eco-driving performance. For example, in scenario "C30\_S20", shown in Figure~\ref{fig:C30S20}, ego-vehicle will start with low acceleration levels to avoid waiting at the intersection since the energy consumption may go much higher if the speed increases rapidly.  Conversely, the HRL will also seeking a high acceleration to improve the time efficiency, if the vehicle can pass the intersection before the end of green phase. For instance, in scenario "C40\_S20", shown in Figure~\ref{fig:C40S20}, the ego-vehicle will also start with a high acceleration to pass the intersection as soon as possible. These time-speed trajectories in Figure~\ref{fig:testing_results} demonstrate the advantage of the HRL method in generating smarter acceleration results than IDM and Graph methods and adjusting the acceleration according to the various traffic environments dynamically.

\begin{table*}[!h]
\centering
\caption{Numerical Testing results for the HRL, Graph methods and the IDM baseline under different traffic scenarios. Improvements for each model compared with the IDM model are shown as Imp. Best Imp. for each scenario is notated by bold style with underlined.}
\label{tab:testing results}
\begin{tabular}{cc|ccccc|ccccc}
\hline
                                          &       & \multicolumn{5}{c|}{Energy consumption {[}J{]}}                                                                            & \multicolumn{5}{c|}{Travel time {[}s{]}}                                                                                   \\ \cline{3-12} 
                                          &       & S10                    & S20                    & S30                    & S40                    & S50                    & S10                    & S20                    & S30                    & S40                    & S50                    \\ \hline
\multicolumn{1}{c|}{\multirow{5}{*}{C0}}  & IDM$\downarrow$   & 1,396.70               & 1,225.89               & 1,093.93               & 1,061.38               & 924.75                 & 71.00                  & 65.50                  & 70.50                  & 70.50                  & 70.50                  \\
\multicolumn{1}{c|}{}                     & Graph$\downarrow$ & 714.02                 & 724.69                 & 760.53                 & 685.18                 & 642.84                 & 76.00                  & 76.00                  & 75.00                  & 75.00                  & 75.00                  \\
\multicolumn{1}{c|}{}                     & Imp.$\uparrow$ & {\ul \textbf{48.88\%}} & 40.88\%                & 30.48\%                & 35.45\%                & 30.49\%                & -7.04\%                & -16.03\%               & -6.38\%                & -6.38\%                & -6.38\%                \\
\multicolumn{1}{c|}{}                     & HRL$\downarrow$   & 751.73                 & 704.47                 & 668.05                 & 627.46                 & 575.51                 & 69.74                  & 63.90                  & 64.14                  & 67.66                  & 63.24                  \\
\multicolumn{1}{c|}{}                     & Imp.$\uparrow$  & 46.18\%                & {\ul \textbf{42.53\%}} & {\ul \textbf{38.93\%}} & {\ul \textbf{40.88\%}} & {\ul \textbf{37.77\%}} & {\ul \textbf{1.77\%}}  & {\ul \textbf{2.44\%}}  & {\ul \textbf{9.02\%}}  & {\ul \textbf{4.03\%}}  & {\ul \textbf{10.30\%}} \\ \hline
\multicolumn{1}{c|}{\multirow{5}{*}{C10}} & IDM$\downarrow$   & 1,366.42               & 1,257.27               & 1,063.54               & 1,030.99               & 894.36                 & 61.00                  & 59.00                  & 60.50                  & 60.50                  & 60.50                  \\
\multicolumn{1}{c|}{}                     & Graph$\downarrow$ & 635.23                 & 697.34                 & 600.06                 & 535.89                 & 538.85                 & 64.00                  & 61.50                  & 63.00                  & 65.00                  & 65.00                  \\
\multicolumn{1}{c|}{}                     & Imp.$\uparrow$  & {\ul \textbf{53.51\%}} & {\ul \textbf{44.54\%}} & {\ul \textbf{43.58\%}} & {\ul \textbf{48.02\%}} & {\ul \textbf{39.75\%}} & {\ul \textbf{-4.92\%}} & -4.24\%                & -4.13\%                & -7.44\%                & -7.44\%                \\
\multicolumn{1}{c|}{}                     & HRL$\downarrow$   & 737.49                 & 699.46                 & 643.57                 & 593.92                 & 599.65                 & 65.36                  & 58.02                  & 62.52                  & 57.18                  & 60.28                  \\
\multicolumn{1}{c|}{}                     & Imp.$\uparrow$  & 46.03\%                & 44.37\%                & 39.49\%                & 42.39\%                & 32.95\%                & -7.15\%                & {\ul \textbf{1.66\%}}  & {\ul \textbf{-3.34\%}} & {\ul \textbf{5.49\%}}  & {\ul \textbf{0.36\%}}  \\ \hline
\multicolumn{1}{c|}{\multirow{5}{*}{C20}} & IDM$\downarrow$   & 1,368.93               & 1,237.88               & 1,033.03               & 1,000.48               & 863.85                 & 53.00                  & 50.00                  & 50.50                  & 50.50                  & 50.50                  \\
\multicolumn{1}{c|}{}                     & Graph$\downarrow$ & 746.47                 & 814.58                 & 663.77                 & 668.81                 & 519.85                 & 49.50                  & 56.00                  & 55.50                  & 54.50                  & 51.50                  \\
\multicolumn{1}{c|}{}                     & Imp.$\uparrow$  & 45.47\%                & 34.20\%                & {\ul \textbf{35.75\%}} & 33.15\%                & {\ul \textbf{39.82\%}} & {\ul \textbf{6.60\%}}  & -12.00\%               & -9.90\%                & -7.92\%                & -1.98\%                \\
\multicolumn{1}{c|}{}                     & HRL$\downarrow$   & 649.09                 & 680.16                 & 678.69                 & 663.96                 & 552.61                 & 51.30                  & 52.30                  & 45.52                  & 49.76                  & 49.30                  \\
\multicolumn{1}{c|}{}                     & Imp.$\uparrow$  & {\ul \textbf{52.58\%}} & {\ul \textbf{45.05\%}} & 34.30\%                & {\ul \textbf{33.64\%}} & 36.03\%                & 3.21\%                 & {\ul \textbf{-4.60\%}} & {\ul \textbf{9.86\%}}  & {\ul \textbf{1.47\%}}  & {\ul \textbf{2.38\%}}  \\ \hline
\multicolumn{1}{c|}{\multirow{5}{*}{C30}} & IDM$\downarrow$   & 1,334.85               & 1,287.27               & 1,047.44               & 1,014.63               & 879.44                 & 49.50                  & 48.00                  & 43.00                  & 43.00                  & 43.00                  \\
\multicolumn{1}{c|}{}                     & Graph$\downarrow$ & 1,038.18               & 900.05                 & 923.07                 & 670.63                 & 632.70                 & 53.00                  & 49.50                  & 50.50                  & 43.00                  & 43.00                  \\
\multicolumn{1}{c|}{}                     & Imp.$\uparrow$  & 22.23\%                & 30.08\%                & 11.87\%                & 33.90\%                & 28.06\%                & -7.07\%                & -3.13\%                & -17.44\%               & 0.00\%                 & 0.00\%                 \\
\multicolumn{1}{c|}{}                     & HRL$\downarrow$   & 716.43                 & 617.21                 & 540.53                 & 541.73                 & 437.68                 & 50.80                  & 45.76                  & 41.78                  & 41.28                  & 37.72                  \\
\multicolumn{1}{c|}{}                     & Imp.$\uparrow$  & {\ul \textbf{46.33\%}} & {\ul \textbf{52.05\%}} & {\ul \textbf{48.40\%}} & {\ul \textbf{46.61\%}} & {\ul \textbf{50.23\%}} & {\ul \textbf{-2.63\%}} & {\ul \textbf{4.67\%}}  & {\ul \textbf{2.84\%}}  & {\ul \textbf{4.00\%}}  & {\ul \textbf{12.28\%}} \\ \hline
\multicolumn{1}{c|}{\multirow{5}{*}{C40}} & IDM$\downarrow$   & 1,405.08               & 1,380.18               & 1,031.00               & 1,004.31               & 869.22                 & 90.50                  & 90.50                  & 42.50                  & 42.50                  & 42.50                  \\
\multicolumn{1}{c|}{}                     & Graph$\downarrow$ & 989.91                 & 1,048.24               & 1,013.90               & 681.27                 & 609.91                 & 87.50                  & 91.50                  & 91.50                  & 43.00                  & 43.00                  \\
\multicolumn{1}{c|}{}                     & Imp.$\uparrow$  & 29.55\%                & 24.05\%                & 1.66\%                 & 32.17\%                & 29.83\%                & 3.31\%                 & -1.10\%                & -115.29\%              & -1.18\%                & -1.18\%                \\
\multicolumn{1}{c|}{}                     & HRL$\downarrow$   & 641.90                 & 632.14                 & 578.50                 & 461.55                 & 436.51                 & 49.28                  & 43.78                  & 40.42                  & 39.10                  & 37.32                  \\
\multicolumn{1}{c|}{}                     & Imp.$\uparrow$  & {\ul \textbf{54.32\%}} & {\ul \textbf{54.20\%}} & {\ul \textbf{43.89\%}} & {\ul \textbf{54.04\%}} & {\ul \textbf{49.78\%}} & {\ul \textbf{45.55\%}} & {\ul \textbf{51.62\%}} & {\ul \textbf{4.89\%}}  & {\ul \textbf{8.00\%}}  & {\ul \textbf{12.19\%}} \\ \hline
\multicolumn{1}{c|}{\multirow{5}{*}{C50}} & IDM$\downarrow$   & 1,380.60               & 1,322.42               & 1,070.02               & 1,037.36               & 900.79                 & 82.00                  & 81.00                  & 77.50                  & 77.50                  & 77.50                  \\
\multicolumn{1}{c|}{}                     & Graph$\downarrow$ & 804.07                 & 958.04                 & 941.17                 & 807.88                 & 770.19                 & 77.50                  & 88.00                  & 81.50                  & 83.50                  & 83.50                  \\
\multicolumn{1}{c|}{}                     & Imp.$\uparrow$  & {\ul \textbf{41.76\%}} & 27.55\%                & 12.04\%                & 22.12\%                & 14.50\%                & {\ul \textbf{5.49\%}}  & -8.64\%                & -5.16\%                & -7.74\%                & -7.74\%                \\
\multicolumn{1}{c|}{}                     & HRL$\downarrow$   & 807.90                 & 786.62                 & 824.45                 & 769.51                 & 697.41                 & 79.76                  & 81.36                  & 81.34                  & 81.32                  & 79.82                  \\
\multicolumn{1}{c|}{}                     & Imp.$\uparrow$  & 41.48\%                & {\ul \textbf{40.52\%}} & {\ul \textbf{22.95\%}} & {\ul \textbf{25.82\%}} & {\ul \textbf{22.58\%}} & 2.73\%                 & {\ul \textbf{-0.44\%}} & {\ul \textbf{-4.95\%}} & {\ul \textbf{-4.93\%}} & {\ul \textbf{-2.99\%}} \\ \hline
\end{tabular}
\end{table*}

\textbf{(2) The Target Speed}: for IDM model, the instantaneous target speed is always the speed limit. For the Graph model, the target speed can be adjusted with respect to the current SPaT information. However, compared with IDM and Graph, HRL model can dynamically control its target speed according to the current traffic situation and achieve a better energy-saving performance. For instance, in Figure~\ref{fig:C50S20}, under the scenario in which the vehicle have to wait for a long red phase, Graph method would like to driving in a slow mode to avoid stop at the intersection, while the HRL will still keep a high speed to approach the intersection. Since we have other vehicles in the traffic, the Graph-based vehicle will approach the stop line behind a long queue and its idea speed trajectory will also be disturbed by these front vehicles, which causes this non-stopping-oriented  speed trajectory planning model performing less than HRL in both energy consumption and travel time. 

\begin{figure}[!h]
    \centering
    \includegraphics[width=0.48\textwidth]{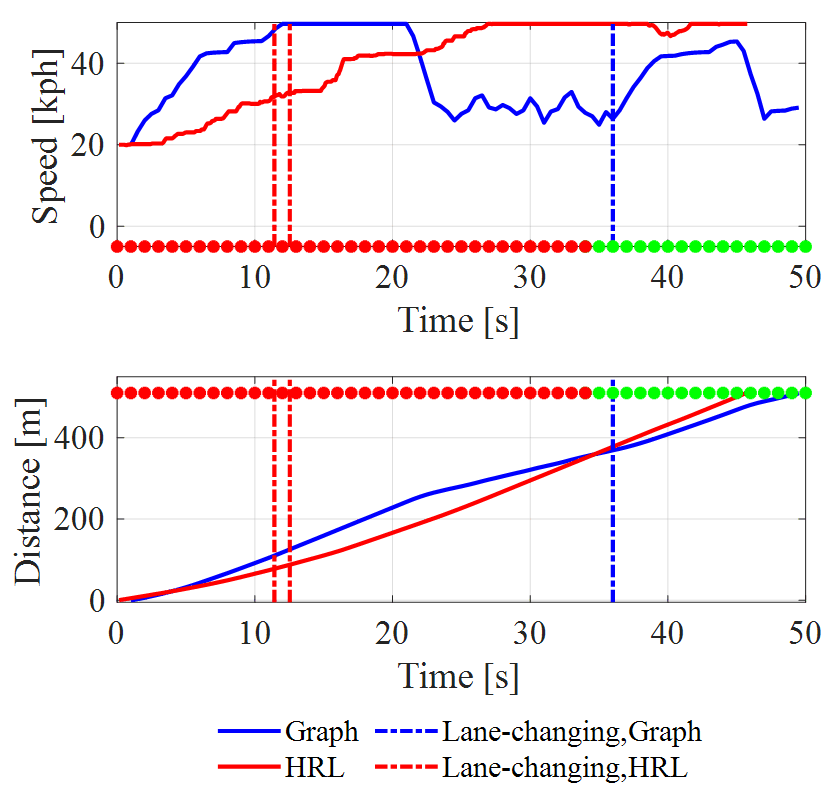}
    \caption{Lane-changing behavior of Graph and HRL.}
    \label{fig:lanechange}
\end{figure}

\textbf{(3) The interaction with the mixed traffic}: original IDM and Graph model can only longitudinally interact with other vehicles. To make a fair comparison, we implemented a rule-based lane-changing model to IDM and Graph for the lateral actions. Although endowed with lateral actions, HRL can interact with surrounding vehicles with much smarter manners, which can be illustrated in Figure~\ref{fig:lanechange}. Figure~\ref{fig:lanechange} shows the time-speed and time-distance diagram for Graph model and HRL model. For both model, there is a slow vehicle driving in front of the ego-vehicle. For Graph model, the vehicle will be impeded to slow down and then wait for several seconds and then make the lane-changing action to pursue the travel time. Nevertheless, for HRL model, since it can perceive the environment via camera, the lane-changing actions happened before approaching the slow front vehicle, which can extremely improve the travel time performance. 

Table \ref{tab:testing results} shows the numerical testing results for the proposed HRL method, model-based Graph method, and the baseline IDM. From Table \ref{tab:testing results}, in all 30 scenarios, our model can achieve the best energy-consumption improvement in 21 scenarios and best travel-time improvement in 27 scenarios. 

\begin{table}[!h]
\centering
\caption{Average performance of different eco-driving methods (Imp. represents the improvement for HRL compared with Graph method).}
\begin{tabular}{c|ccc|ccc}
\hline
     & \multicolumn{3}{c|}{Energy consumtion {[}J{]}}   & \multicolumn{3}{c|}{Travel time {[}s{]}} \\ \cline{2-7} 
     & Graph   & HRL     & \multicolumn{1}{c|}{Imp.}    & Graph        & HRL         & Imp.        \\ \hline
C0   & 37.23\% & 41.26\% & 5.71\%                       & -8.44\%      & 5.51\%      & 12.82\%     \\
C10   & 45.88\% & 41.05\% & -9.15\%                      & -5.63\%      & -0.59\%     & 4.72\%      \\
C20   & 37.68\% & 40.32\% & 4.34\%                       & -5.04\%      & 2.46\%      & 6.78\%      \\
C30   & 25.23\% & 48.72\% & 30.78\%                      & -5.53\%      & 4.23\%      & 9.05\%      \\
C40   & 23.45\% & 51.25\% & 35.69\%                      & -23.09\%     & 24.45\%     & 34.79\%     \\
C50   & 23.59\% & 30.67\% & 8.80\%                       & -4.76\%      & -2.12\%     & 2.37\%      \\
Avg. & 32.18\% & 42.21\% & \multicolumn{1}{c|}{12.70\%} & -8.75\%      & 5.66\%      & 11.75\%     \\ \hline
\end{tabular}
\label{table:test_average}
\end{table}
Table \ref{table:test_average} shows the average performance of different eco-driving methods under different initial signal-phase condition. From the table, compared with Graph model, the HRL method can save 12.70\% (up to 35.69\%) energy consumption and 11.75\% (up to 34.79\%) travel time on average per journey in all testing scenarios. 
To decouple the possible impact by ignoring the brake charging, we also evaluated all those models by adding the regen-braking charging. The results showed that the energy consumption decreased for all methods with regen braking charging. Specifically, with regen-braking energy, HRL method can reduce 23.60\% and 41.59\% energy consumption compared with Graph and IDM. The overall relative performance among methods is still the same, which demonstrate the advantages of utilizing learning-based methods. 
\begin{figure}[!h]
    \centering
    \includegraphics[width=0.42\textwidth]{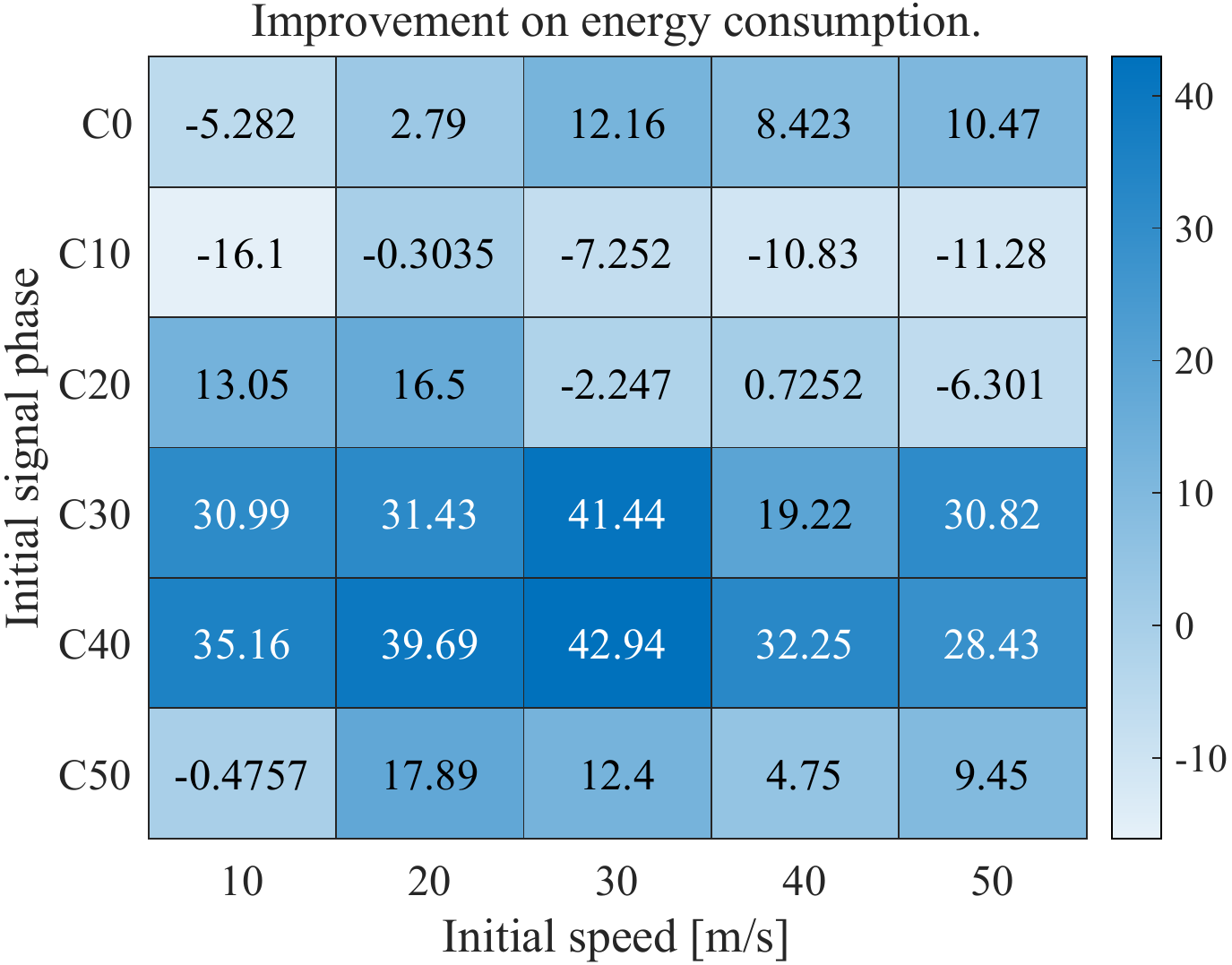}
    \caption{The heat map of improvement on energy consumption.}
    \label{fig:energy_heatmap}
\end{figure}

\begin{figure}[!h]
    \centering
    \includegraphics[width=0.42\textwidth]{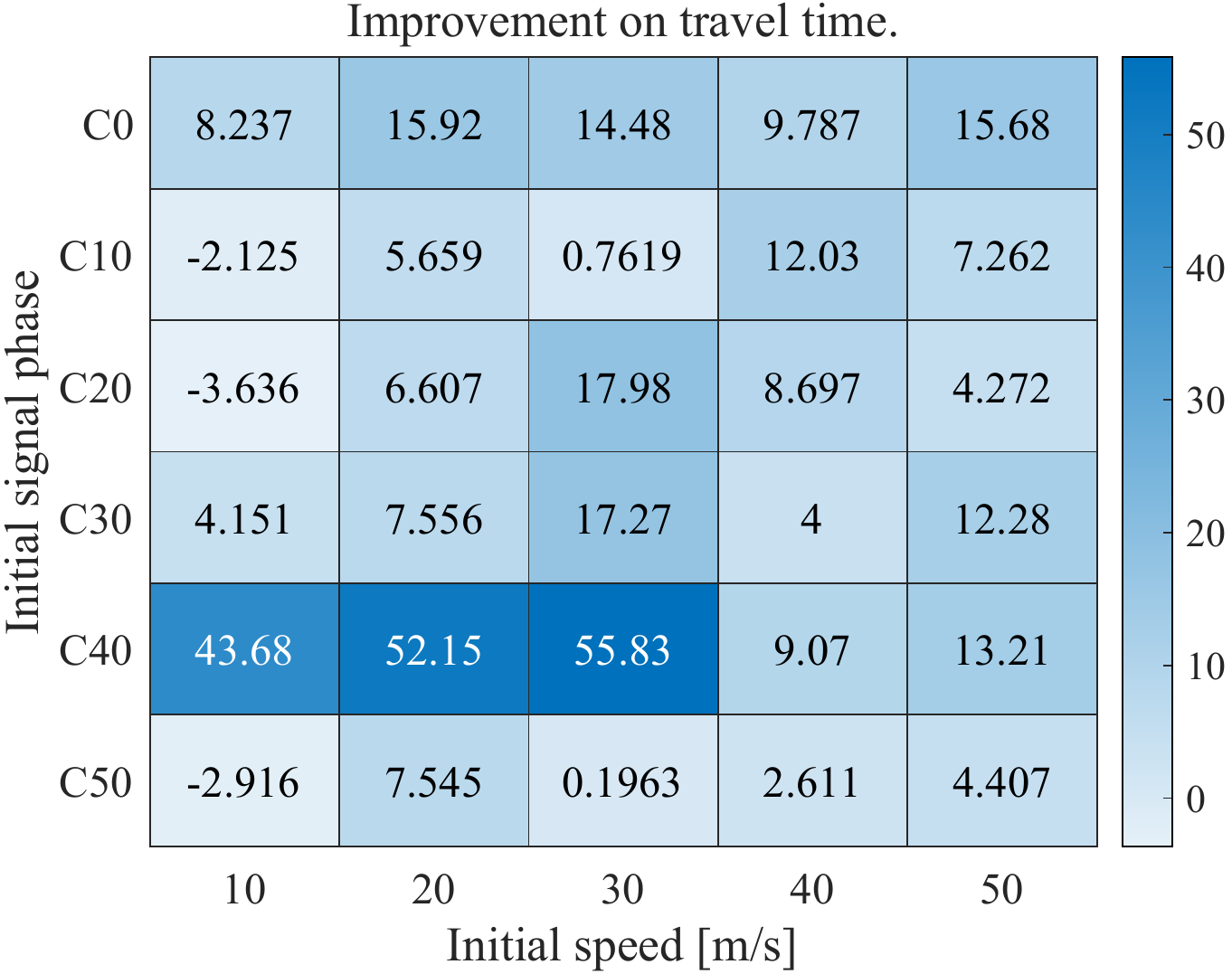}
    \caption{The heat map of improvement on travel time.}
    \label{fig:time_heatmap}
\end{figure}
Furthermore, Figure~\ref{fig:energy_heatmap} and Figure~\ref{fig:time_heatmap} demonstrates the deeper connections behind the vehicle entry speed and entry time, and shows that most of the dark areas (in which the Eco-Driving model has better energy-saving/travel time performance) gather together. On the contrary, for the upper right area, the performance of HRL is relatively low. According to the analysis, the entry time and entry speed will cooperatively influence the performance of HRL method, which reminds us that there is a sweet spot for the HRL method. If we can control the vehicle to enter the intersection with a proper zone, the HRL-based eco-driving approach will get its best performance. For example, when the entry time is C50 and entry speed of the vehicle is 30 $kph$, the HRL method has the best improvement. In this situation, comparing with Graph method, HRL method saves 42.94\% energy consumption and 55.83\% travel time, respectively, which is a huge improvement compared with Graph method.

\section{CONCLUSIONS \& FUTURE WORK}
In this paper, we proposed an eco-driving approach for CAVs under mixed intersection traffic by designing a hybrid reinforcement learning (HRL) framework. The vehicle activity data from On-Board Diagnosis (OBD), front camera, on-board radar and the V2I communication are collected by the HRL framework. The output of HRL is longitudinal acceleration value and lateraltarget lane. The HRL framework combines a rule-based strategy and a deep reinforcement learning-based policy learning algorithm. The predefined rules are designed on the basis of different driving conditions to ensure the driving safety and efficiency. The eco-driving RL algorithm consists of a spatiotemporal data preprocessor, a feature extracting network, and a policy network. The HRL learns the optimal eco-driving strategy through Long-Short Term Reward (LSTR) model, with both short-term and long-term benefits. Numerical experiments are conducted in Unity at a signalized intersection with the mixed traffic situation and a state-of-art model-based Eco-Driving method is implemented to assess the performance of the proposed method.

According to the experiments, the proposed HRL-based ego-vehicle can traverse through a signalized intersection with eco-driving strategy under mixed traffic conditions. The HRL method can reduce the energy consumption by 12.70\%/56.01\% comparing with Graph/IDM methods and can save 11.75\%/5.66\% in travel time compared with Graph/IDM methods. The proposed framework can also be readily implemented to other types of vehicles by replacing the energy-reward function and vehicle dynamic model. 

For future work, the performance of the different types of vehicles (e.g. heavy-duty trucks) will be conducted and analyzed. In addition, cooperative eco-driving strategy will be conducted by applying multi-vehicle agents in complex traffic network. Furthermore, more experiments including micro-simulation and field experiments will be conducted to analyze the performance in more realistic situations.

\section*{Acknowledgment}

This research is supported by The National Center for Sustainable Transportation (NCST).




\bibliographystyle{IEEEtran}
\bibliography{references}{}

\begin{thebibliography}{10}
\providecommand{\url}[1]{#1}
\csname url@samestyle\endcsname
\providecommand{\newblock}{\relax}
\providecommand{\bibinfo}[2]{#2}
\providecommand{\BIBentrySTDinterwordspacing}{\spaceskip=0pt\relax}
\providecommand{\BIBentryALTinterwordstretchfactor}{4}
\providecommand{\BIBentryALTinterwordspacing}{\spaceskip=\fontdimen2\font plus
\BIBentryALTinterwordstretchfactor\fontdimen3\font minus
  \fontdimen4\font\relax}
\providecommand{\BIBforeignlanguage}[2]{{%
\expandafter\ifx\csname l@#1\endcsname\relax
\typeout{** WARNING: IEEEtran.bst: No hyphenation pattern has been}%
\typeout{** loaded for the language `#1'. Using the pattern for}%
\typeout{** the default language instead.}%
\else
\language=\csname l@#1\endcsname
\fi
#2}}
\providecommand{\BIBdecl}{\relax}
\BIBdecl

\bibitem{vahidi2018energy}
A.~Vahidi and A.~Sciarretta, ``Energy saving potentials of connected and
  automated vehicles,'' \emph{Transportation Research Part C: Emerging
  Technologies}, vol.~95, pp. 822--843, 2018.

\bibitem{fagnant2015preparing}
D.~J. Fagnant and K.~Kockelman, ``Preparing a nation for autonomous vehicles:
  opportunities, barriers and policy recommendations,'' \emph{Transportation
  Research Part A: Policy and Practice}, vol.~77, pp. 167--181, 2015.

\bibitem{INRIX}
INRIX, \emph{Los Angeles Tops INRIX Global Congestion Ranking,
  http://inrix.com/press-releases/scorecard-2017/}, accessed July 10, 2019.

\bibitem{GGE}
\emph{National Greenhouse Gas Emissions Data Report}, U.S. Environ. Protection
  Agency, Washington, DC, 2013.

\bibitem{rios2016survey}
J.~Rios-Torres and A.~A. Malikopoulos, ``A survey on the coordination of
  connected and automated vehicles at intersections and merging at highway
  on-ramps,'' \emph{IEEE Transactions on Intelligent Transportation Systems},
  vol.~18, no.~5, pp. 1066--1077, 2016.

\bibitem{lee2012development}
J.~Lee and B.~Park, ``Development and evaluation of a cooperative vehicle
  intersection control algorithm under the connected vehicles environment,''
  \emph{IEEE Transactions on Intelligent Transportation Systems}, vol.~13,
  no.~1, pp. 81--90, 2012.

\bibitem{guler2014using}
S.~I. Guler, M.~Menendez, and L.~Meier, ``Using connected vehicle technology to
  improve the efficiency of intersections,'' \emph{Transportation Research Part
  C: Emerging Technologies}, vol.~46, pp. 121--131, 2014.

\bibitem{elhenawy2015intersection}
M.~Elhenawy, A.~A. Elbery, A.~A. Hassan, and H.~A. Rakha, ``An intersection
  game-theory-based traffic control algorithm in a connected vehicle
  environment,'' in \emph{2015 IEEE 18th International Conference on
  Intelligent Transportation Systems}.\hskip 1em plus 0.5em minus 0.4em\relax
  IEEE, 2015, pp. 343--347.

\bibitem{zhang2016optimal}
Y.~J. Zhang, A.~A. Malikopoulos, and C.~G. Cassandras, ``Optimal control and
  coordination of connected and automated vehicles at urban traffic
  intersections,'' in \emph{2016 American Control Conference (ACC)}.\hskip 1em
  plus 0.5em minus 0.4em\relax IEEE, 2016, pp. 6227--6232.

\bibitem{hao2018eco}
P.~Hao, G.~Wu, K.~Boriboonsomsin, and M.~J. Barth, ``Eco-approach and departure
  (ead) application for actuated signals in real-world traffic,'' \emph{IEEE
  Transactions on Intelligent Transportation Systems}, vol.~20, no.~1, pp.
  30--40, 2018.

\bibitem{ye2018prediction}
F.~Ye, P.~Hao, X.~Qi, G.~Wu, K.~Boriboonsomsin, and M.~J. Barth,
  ``Prediction-based eco-approach and departure at signalized intersections
  with speed forecasting on preceding vehicles,'' \emph{IEEE Transactions on
  Intelligent Transportation Systems}, vol.~20, no.~4, pp. 1378--1389, 2018.

\bibitem{wang2019cooperative}
Z.~Wang, G.~Wu, and M.~J. Barth, ``Cooperative eco-driving at signalized
  intersections in a partially connected and automated vehicle environment,''
  \emph{IEEE Transactions on Intelligent Transportation Systems}, vol.~21,
  no.~5, pp. 2029--2038, 2019.

\bibitem{de2016eco}
G.~De~Nunzio, C.~C. De~Wit, P.~Moulin, and D.~Di~Domenico, ``Eco-driving in
  urban traffic networks using traffic signals information,''
  \emph{International Journal of Robust and Nonlinear Control}, vol.~26, no.~6,
  pp. 1307--1324, 2016.

\bibitem{xia2012field}
H.~Xia, K.~Boriboonsomsin, F.~Schweizer, A.~Winckler, K.~Zhou, W.-B. Zhang, and
  M.~Barth, ``Field operational testing of eco-approach technology at a
  fixed-time signalized intersection,'' in \emph{2012 15th International IEEE
  Conference on Intelligent Transportation Systems}.\hskip 1em plus 0.5em minus
  0.4em\relax IEEE, 2012, pp. 188--193.

\bibitem{hao2018connected}
P.~Hao, K.~Boriboonsomsin, C.~Wang, G.~Wu, and M.~Barth, ``Connected
  eco-approach and departure (ead) system for diesel trucks,'' Tech. Rep.,
  2018.

\bibitem{ye2017prediction}
F.~Ye, P.~Hao, X.~Qi, G.~Wu, K.~Boriboonsomsin, and M.~Barth,
  ``Prediction-based eco-approach and departure strategy in congested urban
  traffic,'' 2017.

\bibitem{hu2021cut}
J.~Hu, Z.~Zhang, L.~Xiong, H.~Wang, and G.~Wu, ``Cut through traffic to catch
  green light: eco approach with overtaking capability,'' \emph{Transportation
  research part C: emerging technologies}, vol. 123, p. 102927, 2021.

\bibitem{zhao2018platoon}
W.~Zhao, D.~Ngoduy, S.~Shepherd, R.~Liu, and M.~Papageorgiou, ``A platoon based
  cooperative eco-driving model for mixed automated and human-driven vehicles
  at a signalised intersection,'' \emph{Transportation Research Part C:
  Emerging Technologies}, vol.~95, pp. 802--821, 2018.

\bibitem{huang2018eco}
Y.~Huang, E.~C. Ng, J.~L. Zhou, N.~C. Surawski, E.~F. Chan, and G.~Hong,
  ``Eco-driving technology for sustainable road transport: A review,''
  \emph{Renewable and Sustainable Energy Reviews}, vol.~93, pp. 596--609, 2018.

\bibitem{jordan2015machine}
M.~I. Jordan and T.~M. Mitchell, ``Machine learning: Trends, perspectives, and
  prospects,'' \emph{Science}, vol. 349, no. 6245, pp. 255--260, 2015.

\bibitem{lecun2015deep}
Y.~LeCun, Y.~Bengio, and G.~Hinton, ``Deep learning,'' \emph{nature}, vol. 521,
  no. 7553, p. 436, 2015.

\bibitem{sutton2018reinforcement}
R.~S. Sutton and A.~G. Barto, \emph{Reinforcement learning: An
  introduction}.\hskip 1em plus 0.5em minus 0.4em\relax MIT press, 2018.

\bibitem{Silver2016Mastering}
D.~Silver, A.~Huang, C.~J. Maddison, A.~Guez, L.~Sifre, G.~V.~D. Driessche,
  J.~Schrittwieser, I.~Antonoglou, V.~Panneershelvam, and M.~Lanctot,
  ``Mastering the game of go with deep neural networks and tree search,''
  \emph{Nature}, vol. 529, no. 7587, pp. 484--489, 2016.

\bibitem{Mnih2015Human}
V.~Mnih, K.~Kavukcuoglu, D.~Silver, A.~A. Rusu, J.~Veness, M.~G. Bellemare,
  A.~Graves, M.~Riedmiller, A.~K. Fidjeland, and G.~Ostrovski, ``Human-level
  control through deep reinforcement learning,'' \emph{Nature}, vol. 518, no.
  7540, p. 529, 2015.

\bibitem{sallab2017deep}
A.~E. Sallab, M.~Abdou, E.~Perot, and S.~Yogamani, ``Deep reinforcement
  learning framework for autonomous driving,'' \emph{Electronic Imaging}, vol.
  2017, no.~19, pp. 70--76, 2017.

\bibitem{desjardins2011cooperative}
C.~Desjardins and B.~Chaib-Draa, ``Cooperative adaptive cruise control: A
  reinforcement learning approach,'' \emph{IEEE Transactions on intelligent
  transportation systems}, vol.~12, no.~4, pp. 1248--1260, 2011.

\bibitem{shalev2016safe}
S.~Shalev-Shwartz, S.~Shammah, and A.~Shashua, ``Safe, multi-agent,
  reinforcement learning for autonomous driving,'' \emph{arXiv preprint
  arXiv:1610.03295}, 2016.

\bibitem{chen2018deep}
J.~Chen, Z.~Wang, and M.~Tomizuka, ``Deep hierarchical reinforcement learning
  for autonomous driving with distinct behaviors,'' in \emph{2018 IEEE
  Intelligent Vehicles Symposium (IV)}.\hskip 1em plus 0.5em minus 0.4em\relax
  IEEE, 2018, pp. 1239--1244.

\bibitem{1998Reinforcement}
R.~Sutton and A.~Barto, \emph{Reinforcement Learning:An Introduction}.\hskip
  1em plus 0.5em minus 0.4em\relax MIT Press, 1998.

\bibitem{bertsekas1995dynamic}
D.~P. Bertsekas, D.~P. Bertsekas, D.~P. Bertsekas, and D.~P. Bertsekas,
  \emph{Dynamic programming and optimal control}.\hskip 1em plus 0.5em minus
  0.4em\relax Athena scientific Belmont, MA, 1995, vol.~1, no.~2.

\bibitem{treiber2000congested}
M.~Treiber, A.~Hennecke, and D.~Helbing, ``Congested traffic states in
  empirical observations and microscopic simulations,'' \emph{Physical review
  E}, vol.~62, no.~2, p. 1805, 2000.

\bibitem{ye2016hybrid}
F.~Ye, G.~Wu, K.~Boriboonsomsin, and M.~J. Barth, ``A hybrid approach to
  estimating electric vehicle energy consumption for ecodriving applications,''
  in \emph{2016 IEEE 19th International Conference on Intelligent
  Transportation Systems (ITSC)}.\hskip 1em plus 0.5em minus 0.4em\relax IEEE,
  2016, pp. 719--724.

\bibitem{Schaul2015Prioritized}
T.~Schaul, J.~Quan, I.~Antonoglou, and D.~Silver, ``Prioritized experience
  replay,'' \emph{Computer Science}, 2015.

\bibitem{wang2015dueling}
Z.~Wang, T.~Schaul, M.~Hessel, H.~Van~Hasselt, M.~Lanctot, and N.~De~Freitas,
  ``Dueling network architectures for deep reinforcement learning,'' pp.
  1995--2003, 2015.

\bibitem{albawi2017understanding}
S.~Albawi, T.~A. Mohammed, and S.~Al-Zawi, ``Understanding of a convolutional
  neural network,'' in \emph{2017 International Conference on Engineering and
  Technology (ICET)}.\hskip 1em plus 0.5em minus 0.4em\relax Ieee, 2017, pp.
  1--6.

\bibitem{hochreiter1997long}
S.~Hochreiter and J.~Schmidhuber, ``Long short-term memory,'' \emph{Neural
  computation}, vol.~9, no.~8, pp. 1735--1780, 1997.

\bibitem{Kingma2014Adam}
D.~P. Kingma and J.~Ba, ``Adam: A method for stochastic optimization,''
  \emph{arXiv preprint arXiv:1412.6980}, 2014.

\bibitem{juliani2018unity}
A.~Juliani, V.-P. Berges, E.~Teng, A.~Cohen, J.~Harper, C.~Elion, C.~Goy,
  Y.~Gao, H.~Henry, M.~Mattar \emph{et~al.}, ``Unity: A general platform for
  intelligent agents,'' \emph{arXiv preprint arXiv:1809.02627}, 2018.

\bibitem{bai2019deep}
Z.~Bai, B.~Cai, W.~Shangguan, and L.~Chai, ``Deep reinforcement learning based
  high-level driving behavior decision-making model in heterogeneous traffic,''
  \emph{arXiv preprint arXiv:1902.05772}, 2019.

\bibitem{min2018deep}
K.~Min and H.~Kim, ``Deep q learning based high level driving policy
  determination,'' in \emph{2018 IEEE Intelligent Vehicles Symposium
  (IV)}.\hskip 1em plus 0.5em minus 0.4em\relax IEEE, 2018, pp. 226--231.

\bibitem{abadi2016tensorflow}
M.~Abadi, P.~Barham, J.~Chen, Z.~Chen, A.~Davis, J.~Dean, M.~Devin,
  S.~Ghemawat, G.~Irving, M.~Isard \emph{et~al.}, ``Tensorflow: A system for
  large-scale machine learning,'' in \emph{12th $\{$USENIX$\}$ symposium on
  operating systems design and implementation ($\{$OSDI$\}$ 16)}, 2016, pp.
  265--283.

\end{thebibliography}
%



%
\begin{IEEEbiography}
    [{\includegraphics[width=1in,height=1.25in,clip,keepaspectratio]{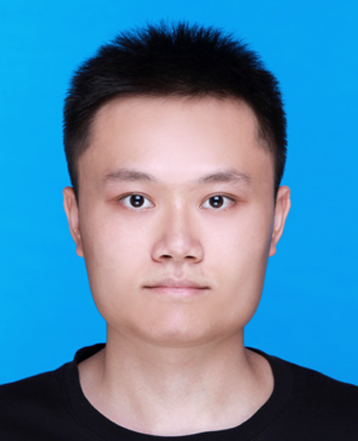}}]{Zhengwei Bai}
(M'21) received the B.E. and M.S. degrees from Beijing Jiaotong University, Beijing, China, in 2017 and 2020, respectively. He is currently a Ph.D. student in electrical and computer engineering at the University of California at Riverside. His research focuses on perception and control for connected driving automation (CDA).
\end{IEEEbiography}

\begin{IEEEbiography}
    [{\includegraphics[width=1in,height=1.25in,clip,keepaspectratio]{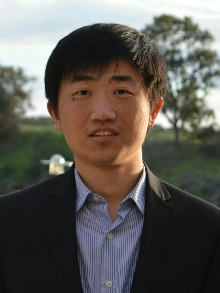}}]{Peng Hao}
(M'16) received Ph.D. degree in transportation engineering from Rensselaer Polytechnic Institute in 2013. He is currently an Assistant Research Engineer with the Transportation Systems Research Group, Center for Environmental Research and Technology, Bourns College of Engineering, University of California, Riverside, USA. His research interests include connected vehicles, eco-approach and departure, sensor-aided modeling, signal control and traffic operations.
\end{IEEEbiography}

\begin{IEEEbiography}
    [{\includegraphics[width=1in,height=1.25in,clip,keepaspectratio]{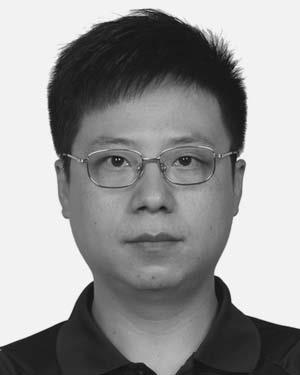}}]{Wei Shangguan}
(M’14) received the B.S, M.S., and Ph.D. degrees from Harbin Engineering University, Harbin, China, in2002, 2005, and 2008, respectively. From 2008 to 2011, he was a Lecturer with the School of Electronic and Information Engineering, Beijing Jiaotong University, Beijing, China. From 2013 to 2014, he was an Academic Visitor with the University College London, London, U.K. He is currently an Associate Professor and a Supervisor of Master with Beijing Jiaotong University. His professional research interests include train control system (Chinese Train Control System (CTCS)/ Europe Train Control System (ETCS)/ The European Railway Traffic Management System (ERTMS)), system modeling, simulation and testing, GNSS (GPS, Galileo, Glonass, and Beidou Navigation Satellite System (BDS))/Geographic Information System (GIS), integrated navigation, intelligent transportation system, and Cooperative Vehicle Infrastructure System of China (CVIS-C).
\end{IEEEbiography}

\begin{IEEEbiography}
    [{\includegraphics[width=1in,height=1.25in,clip,keepaspectratio]{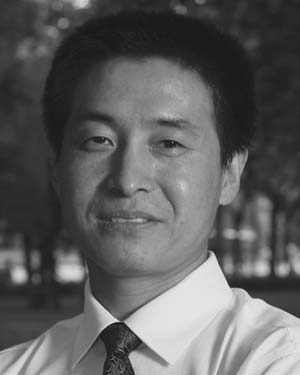}}]{Baigen Cai}
(SM’13) received the B.S., M.S., and Ph.D. degrees from Beijing Jiaotong University, Beijing, China, in 1987, 1990, and 2010, respectively, all in traffic information engineering and control. From 1998 to 1999, he was a Visiting Scholar with Ohio State University. Since 1990, he has been on the Faculty of the School of Electronic and Information Engineering, Beijing Jiaotong University, where he is currently a Professor and the Chief of the Science and Technology Division. His research interests include train control system, intelligent transportation system, GNSS navigation, multi-sensor fusion, and intelligent traffic control.
\end{IEEEbiography}

\begin{IEEEbiography}
    [{\includegraphics[width=1in,height=1.25in,clip,keepaspectratio]{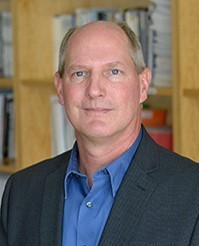}}]
    {Matthew J. Barth}
(M'90-SM'00-F'14) received the M.S. and Ph.D degree in electrical and computer engineering from the University of California at Santa Barbara, in 1985 and 1990, respectively. He is currently the Yeager Families Professor with the College of Engineering, University of California at Riverside, USA. He is also serving as the Director for the Center for Environmental Research and Technology. His current research interests include ITS and the environment, transportation/emissions modeling, vehicle activity analysis, advanced navigation techniques, electric vehicle technology, and advanced sensing and control. Dr. Barth has been active in the IEEE Intelligent Transportation System Society for many years, serving as a Senior Editor for both the Transactions of ITS and the Transactions on Intelligent Vehicles. He served as the IEEE ITSS President for 2014 and 2015 and is currently the IEEE ITSS Vice President of Education.
\end{IEEEbiography}







\end{document}